\documentclass[aps,prl,twocolumn,superscriptaddress,nofootinbib]{revtex4-2}
\usepackage{graphicx,latexsym, color,dsfont}
\usepackage{amsmath,amsfonts,amsbsy,amssymb,amsthm}
\usepackage[utf8]{inputenc}

\usepackage{listings}
\usepackage[mathscr]{euscript}
\usepackage[T1]{fontenc}
\usepackage{mathtools,physics}
\usepackage[english]{babel}
\usepackage{scalerel}
\usepackage{tikz}

\usepackage{bm}

\usepackage[normalem]{ulem}
\usepackage{tabularx}
\usepackage{booktabs}
\usepackage{comment}
\usepackage[hmargin=.7in,vmargin=1.1in]{geometry}
\usepackage[bookmarksnumbered=true,bookmarksopen=true]{hyperref}
\hypersetup{colorlinks,
	linkcolor=[rgb]{0,0.2,0.6},
	citecolor=[rgb]{0,0.2,0.6}}
 \usepackage[capitalise]{cleveref}

\usepackage[all,color]{xy}
\usepackage{qcircuit}
\usepackage{braket}
\usepackage{ragged2e}

\newcommand{\be}{\begin{eqnarray}}
\newcommand{\ee}{\end{eqnarray}}

\newcommand{\ex}{\text{e} }

\newcommand{\spc}[1]{\mathcal{#1}}

\def\>{\rangle}
\def\<{\langle}

\newcommand{\map}[1]{\mathcal{#1}}

\newtheorem{theo}{Theorem}

\newcommand{\tH}{\mathcal{H}}
\newcommand{\tI}{\mathcal{I}}

\newcommand{\tV}{\mathcal{V}}

\renewcommand{\<}{\langle}
\renewcommand{\>}{\rangle}

\begin{document}

\title{A Quantum Superposition of Black Hole Evaporation Histories: Recovering Unitarity}

\author{Ali Akil}
\affiliation{%
Department of Computer Science, The University of Hong Kong, Hong Kong Island, Hong Kong S.A.R. China
}

\author{Lorenzo Giannelli}

\affiliation{%
Department of Computer Science, The University of Hong Kong, Hong Kong Island, Hong Kong S.A.R. China
}

\author{Leonardo Modesto}
\affiliation{Dipartimento di Fisica, Universit\`a di Cagliari, Cittadella Universitaria, 09042 Monserrato, Italy}
 \affiliation{INFN, Sezione di Cagliari, Cittadella Universitaria, 09042 Monserrato, Italy
 \email{leonardo.modesto@unica.it}
 }

\author{Oscar Dahlsten}
\email{oscar.dahlsten@cityu.edu.hk}
\affiliation{Department of Physics, City University of Hong Kong, Tat Chee Avenue, Kowloon, Hong Kong SAR}
\affiliation{Institute of Nanoscience and Applications, Southern University of Science and Technology, Shenzhen 518055, China}

\author{Giulio Chiribella}
\email{giulio@hku.hk}
\affiliation{%
Department of Computer Science, The University of Hong Kong, Hong Kong Island, Hong Kong S.A.R. China
}

\author{Caslav Brukner}

\affiliation{%
\email{caslav.brukner@univie.ac.at}
	Vienna Center for Quantum Science and Technology (VCQ), Faculty of Physics, University of Vienna, Boltzmanngasse 5, A-1090 Vienna, Austria
}%
\affiliation{%
	Institute of Quantum Optics and Quantum Information (IQOQI), Austrian Academy of Sciences, Boltzmanngasse 3, A-1090 Vienna, Austria
}%

\begin{abstract}
Black hole evaporation is one of the most striking phenomena at the interface between gravity and quantum physics. In Hawking's semi-classical treatment, where matter is quantum mechanical and the spacetime is definite and classical, evaporation leads to an apparent loss of unitarity of the overall evolution, and to the so-called black hole information paradox. Here, we go beyond this semi-classical treatment and formulate a toy quantum model of black hole evaporation that allows the black hole to evolve into a superposition of being fully evaporated and not fully evaporated, consistent with the Hawking particles being in a coherent superposition of different energy levels. We model Hawking particle production by the repeated action of quantum-controlled unitaries, generating emission from the quantum black hole and accounting for a quantum coherent back-reaction on the black hole matter state. We show that the probability of full annihilation of the black hole matter increases with time until the black hole is, asymptotically, fully evaporated in every branch of the quantum superposition. We prove that under natural assumptions, this evaporation model is unitary, such that the initial state can in principle be recovered from the final asymptotic state of the radiation. 
\end{abstract}

\maketitle

\noindent

\section{Introduction} 
Black hole evaporation~\cite{Hawking74,Hawking76}  is a fundamental testbed for our understanding of the interface between gravity and quantum physics. When quantum field theory is applied on a black hole background, it predicts a positive energy flux directed outwards from the horizon, and a negative energy flux directed towards the interior of the black hole~\cite{BirrelDavies,Fabbri}. The systems carrying these fluxes exhibit quantum correlations, and are typically modelled as 2-mode squeezed states~\cite{Fabbri}.
The negative flux into the interior causes the black hole to gradually lose mass,
until it fully evaporates or other effects (potentially of Quantum Gravity origin) intervene. The evaporation process can be interpreted as resulting from the creation of particle pairs near the horizon, the excitations of the 2-mode squeezed states~\cite{Fabbri}, with particles created inside having negative energy~\cite{NegativeE}.
A major conceptual problem, known as the black hole information paradox, arises from the fact that the state of the radiation computed in the standard semiclassical setting (definite classical background with quantum matter) seems to be independent of the state of the matter forming the black hole~\cite{HawkingInfo}. Upon evaporation, the information about the initial state appears to be lost, in contradiction with the postulate that isolated quantum systems evolve unitarily. 

The black hole information paradox has stimulated valuable research on a range of topics, including the typical behaviour of the entropy of large quantum systems~\cite{Page,Page2,Page0}, the modelling of black holes and their radiation as quantum circuits~\cite{Hayden,Lloyd,Harlow,ChrisAkers}, the study of post-evaporation black hole remnants in modified and quantum gravity~\cite{Adler:2001vs,Stable,Rovelli2,Perez,Rovelli} and the thermodynamics of regular (non-singular) black holes~\cite{Bambi:2023try, Leonardo23, Bambi:2016yne, Akil2}. Debates were sparked between ideas like measurements on the inside and outside of a black hole being complementary~\cite{Susskind1,Susskind2,Complementarity} and high energy firewall surfaces at the event horizon~\cite{Braunstein,AMPS}, shedding light on different angles of the paradox, including the relation with quantum cloning~\cite{AMPS} and with violations of the equivalence principle~\cite{Braunstein:2014nwa}. Other insightful approaches include Refs.~\cite{Oscar,Mathur,Penington,Penington:2019kki,Almheiri:2019qdq,Maldacena,Ours,Giddings, Chen:2021jzx,Perez:2023ctt, Raju:2024gvc,Tokusumi,Adami,Hausmann:2025jyx,Akil:2025reh,bhinfobook}. Going beyond theoretical investigations, analogous systems present a way to emulate real experiments on black holes~\cite{Carusotto:2008ep,Balbinot:2007de,Ma:2021xqf,AnaBHEL:2022sri,Lin:2021bpe,Recati:2009ya}.  Ref~\cite{Ours} provided an evaporation model with the black hole mass as a quantum degree of freedom, entangled with the Hawking radiation, but stopped short of providing a unitary model. Finally, quantum superpositions of black hole masses have also been considered in the context of decoherence near black holes in Refs. \cite{HowlAkil,Zych1,Zych2}.

In this article, we present a quantum model for black hole evaporation. A key novelty of our approach is that the black hole is not treated semiclassically—i.e., with quantum matter fields evolving on a definite, classical spacetime background. Instead, we consider the black hole itself to be in a quantum superposition of mass eigenstates, including both fully evaporated and non-evaporated branches. We assume evaporation happens via negative energy particles inside the black hole event horizon interacting and annihilating with the black hole matter. The superposition of black hole masses then arises because the Hawking radiation is created in a superposition of all allowed energies, and each energy leads to a different amount of black hole matter annihilation.  Radiation emission is a repeated action of a squeezing operator, controlled over the quantum state of the total black hole mass. The repeated action of squeezing and annihilation accounts for the mutual back reactions between the radiation emission and the black hole state, which Hawking noted were missing from his treatment~\cite{Hawking74, Hawking76}. The way we account for back-reaction preserves the quantumness of the black hole matter rather than treating it as a deterministically changing independent classical parameter. In subsequent radiation emissions, every branch of the black hole state radiates at a different rate. We prove analytically that the probability of full evaporation in every branch of the black hole superposition strictly increases over time, asymptotically reaching 1.

In order to further explore what our quantum evaporation model may potentially imply for information recovery from the black hole radiation, we attempt an extension of the single-subsystem black hole into two subsystems, carrying fine-grained quantum information in  the distribution over their \color{black} relative masses and phases. We propose a model for evaporation and annihilation of subsystems by the negative energy flux. 
The combined operation of Hawking pair creation and annihilation is an isometry which corresponds to the action of a unitary on a restricted set of input states. This allows the black hole information to leak out upon evaporation in each branch.
The entanglement across the event horizon disappears asymptotically, in a process that can be interpreted as a quantum coherent variant of entanglement swapping.
The radiation purity is then shown to follow a Page-curve-like shape.  A curious outcome, however, is that some information is recoverable in non-evaporated branches of the black hole superposition. This may possibly not be in contradiction with the impossibility of information leakage in the presence of a horizon in the semi-classical case, since in our approach the black hole is in a quantum superposition, and the horizon is therefore not well-defined in the quantum mechanical sense. However, this may be associated with signalling, in the case where the black hole information is encoded far away from the event horizon. Since our model is not yet rich enough to encode location and propagation dynamics, this begs the need for a more realistic and elaborate treatment. Finally, the overall evaporation process within our model is unitary at all times. Thus, in this quantum model of black hole evaporation, there is no loss of information.

\section{Results}\label{sec:Results}

\noindent \textbf{Quantum evaporation model} \\ 
We now describe the model for the special case of a single system black hole. 
This special case is sufficient to illustrate several key features, including the coherent superposition of different evaporation histories and the purity of the radiation state upon evaporation. However, in order to appreciate the consequences of this model on information preservation, we later address the case of more subsystems. 
   
The initial state is as follows. The black hole mass is a quantum degree of freedom, starting in a mass eigenstate $\ket{M}_{ \rm bh}$ in the Hilbert space $\mathcal{H_{\rm bh}}$.  For simplicity, we assume that $M$ is an integer multiple of a basic unit of mass, denoted by $m_*$. There are also Hawking radiation modes associated with the space outside the black hole. That space is initially empty,  meaning that the corresponding modes are in the vacuum state  $|0\>_{\rm out} :  =  \ket{0}_{\rm out_1}\ket{0}_{\rm out_2} \ldots$, where $\rm out_1,  out_2,\dots$ label the different outside modes. The key Hilbert spaces are therefore $\mathcal H_{\rm bh} \otimes \mathcal H_{\rm out}$. Another auxiliary Hilbert space associated with internal black hole Hawking modes, $\mathcal H_{\rm int}$ will appear only in intermediate steps. We note that our approach assumes a tensor product structure between the interior and exterior degrees of freedom of the black hole. While this is a common starting point in many models, it remains a matter of debate whether such a factorization is physically meaningful. In particular, some proposed resolutions of the firewall paradox suggest that the black hole interior and the late-time radiation may not correspond to independent subsystems, but instead represent different descriptions of the same degrees of freedom \cite{bhinfobook}.

We model the evaporation as a sequence of discrete steps in which the black hole repeatedly emits Hawking radiation. It is common to either model the full radiation as a single two-mode squeezed state, without accounting for back-reaction, or, at most, accounting for the black hole's mass loss as a time-dependent definite mass. As we show in the Methods, the two-mode squeezed state is more accurately understood as describing only one burst of radiation, with the full evaporation process being a sequence of such bursts.

For simplicity of notation, we let each burst of radiation constitute the creation of a single pair of two systems, the elementary excitations underlying the Hawking radiation (see Methods), which we model as
\begin{equation}
\label{eq:Main2}
|\Phi_{m}\>_{\rm int , out} =  c_m\sum_{\omega=0}^{m}   {\rm e}^{-  \pi m \omega }  \ket{-\omega}_{\rm int}  \ket{ \omega} _{\rm out} ,  
\end{equation}  
where $c_m$ is a normalization constant, and $\omega$ is the frequency. The upper bound on the sum reflects the maximum number of quanta that can be emitted, consistent with the total black hole mass. The initial black hole mass is denoted as $M$ and the general instantaneous mass as $m$. The operator that creates $|\Phi_{m}\>_{\rm int , out}$ is referred to as $V_{\rm prod}$, for which $V_{\rm prod}|m\>_{\rm bh}  = |m\>_{\rm bh} |\Phi_m\>_{\rm int, out}$. 

We assume that when a Hawking pair is created, the negative energy system (int) inside the black hole annihilates a part of the black hole matter, reducing the energy of the annihilated black hole matter. The amount of reduction equals $\omega$, the negative energy of the infalling particle. We model that annihilation process by a linear operator, $W$, that acts as  $W \ket{m}_{\rm bh}\ket{-\omega}_{\rm int}\ket{\omega}_{\rm out} = |m-\omega\>_{\rm bh}  \ket{\omega}_{\rm out}$, for every $\omega \le m$. This operator is an isometry when restricted to the physically relevant subspace, where the created pair has equal ``int'' and ``out'' energies. 

We represent the {\em combined} operation of Hawking pair (Eq.\eqref{eq:Main2}) production $V_{\rm prod}$ and the annihilation $W$ via the evaporation operator $V_{\rm ev}$. We view $V_{\rm ev}$ as a single elementary process whose dynamics cannot be decomposed into separate subprocesses $W$ and $V_{\rm prod}$. These are therefore to be understood as effective processes, introduced solely for intuitive visualization and mathematical reasoning. This indivisibility of the process into effective components prohibits any potential agent within the black hole from encoding information in the time interval between pair creation and annihilation, which could otherwise lead to signaling issues (see Methods and Appendix \ref{app:Signalling}). 
$V_{\rm ev}$ is defined, for the case of a single system black hole, by the relations  
\begin{align}\label{Vev}
V_{\rm ev}|m\>_{\rm bh}  = |\Psi_m\>_{\rm bh, out}    \qquad\forall m  \in  \{0, m_*,  \dots,  M\} \, ,
\end{align}
where 
\begin{align}\label{evaporationstep}
 |\Psi_{m}\>_{\rm bh, out}     =  c_m \sum_{\omega=0}^{m}   {\rm e}^{-  \pi m \omega }  |m  -\omega \>_{\rm bh}   \ket{ \omega} _{\rm out}  \, .
\end{align} 
$V_{\rm ev}$ represents one burst of radiation and its effect on the black hole state. The full process of evaporation is modelled by the repeated action of $V_{\rm ev}$.

As we show later, $V_{\rm ev}$ is an isometry, taking sets of orthonormal states to sets of orthonormal states. 
As a consequence of being an isometry, $V_{\rm ev}$ faithfully encodes every quantum superposition of the states $|m\>$  into a quantum superposition of the states $|\Psi_m\>$. The process, as defined in Eq.\eqref{Vev}, is equivalent to the action of a unitary $U$ for which $U\ket{m}\ket{0}_{\rm out}=|\Psi_m\>_{\rm bh, out}$; adding $\ket{0}_{\rm out}$ to all input states makes the dimensions of the input and output states match without altering their inner products.

$V_{\rm ev}$  incorporates a back reaction on both the black hole and the subsequent radiation emissions. $V_{\rm ev}$ creates an amount of radiation that depends on $m$, as can be seen from Eq. \eqref{evaporationstep}. 
From a quantum circuits perspective, $V_{\rm ev}$ is {\em quantum controlled} by $m$.
The amount of reduction of the black hole mass depends on $m$, and any subsequent burst of radiation reflects the updated mass values.

\noindent {\bf The state after successive iterations}\\
The first application of $V_{\rm ev}$ acts on the initial black hole state $\ket{M}_{\rm bh}$ to yield the following joint state of the black hole matter and the radiation: 
 \be\label{evaporationstep2}
V_{\rm ev}\ket{M}_{\rm bh}&=&
|\Psi_{M}^{(1)}\>_{\rm bh, out} \nonumber \\ 
& =&  c_M \sum_{\omega=0}^{M}   {\rm e}^{-  \pi M \omega }  |M  -\omega \>_{\rm bh}   \ket{ \omega} _{\rm out}  \, .
 \ee
Already at this point, the mass of the black hole is entangled with the energy of the Hawking radiation. In other words, the black hole mass has become indefinite as a result of the evaporation process. In the branch of the wavefunction where $\omega  =  M$, the black hole mass has been completely annihilated, and the evaporation process has concluded. From Eq.(\ref{evaporationstep2}), if a measurement of black hole mass or energy is performed, the probability to find the black hole completely evaporated after the first evaporation step is 
   \be p^{(1)}_{\rm ev}  (M) =   |c_M|^{2}   {\rm e}^{-  2\pi M^2}.  \ee
In the branches where the black hole mass has not been fully annihilated yet, the surface gravity at the horizon is non-zero and particle production continues to generate Hawking radiation.

After the second evaporation step, the black hole will be entangled with two outgoing radiation modes, and the composite system will be in the state  
  \begin{align} \label{eq:twostepevap}
\nonumber  |\Psi_{M}^{(2)}\>_{\rm bh, out_1, out_2}   \! \!  = &  c_M \sum_{\omega_1=0}^{M} c_{M-\omega_1} \! \! \sum_{\omega_2=0}^{M-\omega_1}  {\rm e}^{-  \pi M (\omega_1+ \omega_2) }   {\rm e}^{  \pi\omega_1 \omega_2 } \\
   &  \times |M  -\omega_1 -\omega_2\>_{\rm bh}   \ket{ \omega_1} _{\rm out_1}   \ket{ \omega_2} _{\rm out_2}  \, ,       
\end{align}
where $\rm out_1$ and $\rm out_2$ are the outgoing modes generated in the two evaporation steps.  It is instructive to expand the state in Eq.~(\ref{eq:twostepevap}) into the superposition of an evaporated branch and a non-evaporated branch:  \begin{align} \label{eq:bothexplicit}
  &|\Psi_{M}^{(2)}\>_{\rm bh, out_1, out_2}  \nonumber \\ 
& =   c_M \sum_{\omega_1=0}^{M} c_{M-\omega_1} \! \! \sum_{\omega_2=0}^{M-\omega_1}  {\rm e}^{-  \pi M (\omega_1+ \omega_2) }   {\rm e}^{  \pi\omega_1 \omega_2 } \nonumber \\
   &  \hspace{4mm} \times |0\>_{\rm bh}   \ket{ \omega_1} _{\rm out_1}   \ket{ \omega_2} _{\rm out_2}\delta_{M,\omega_1+\omega_2} \nonumber \\
 &+  c_M        \hspace{-1.5mm}\sum_{\omega_1=0}^{M} \hspace{-1.6mm}  c_{M-\omega_1} \hspace{-1mm} \sum_{\omega_2=0}^{M-\omega_1} \hspace{-1mm} {\rm e}^{-  \pi M (\omega_1+ \omega_2) }   {\rm e}^{  \pi\omega_1 \omega_2 }  \nonumber \\
   & \times |M  -\omega_1  -\omega_2\>_{\rm bh}   \ket{ \omega_1} _{\rm out_1}    \! \ket{ \omega_2} _{\rm out_2} \! \! (1- \delta_{M,\omega_1+\omega_2}).
\end{align}
As is evident from this equation, conditioning on the black hole evaporation ($|0\>_{\rm bh}$), the inside and outside systems are disentangled, and the out radiation is in a pure state.

With each step of evaporation, the probability of full annihilation increases. In general, a sequence of $k$ evaporation steps ($k$ applications of the evaporation operator $V_{\rm ev}$ in Eq.(\ref{Vev}))  generates $k$ outgoing radiation modes.  In the Methods, we show that the probability of complete evaporation  after $k$ evaporation steps satisfies the bound
\begin{align}\label{exponential}
p^{(k)}_{\rm ev}  (M)  \ge  1-  \left[ 1 -{\rm e}^{-2\pi M^2}   \left(1-  {\rm e}^{-2\pi Mm_*}\right)  \right]^k\, .
\end{align}  
Hence, the probability of complete evaporation tends to $1$ as the number of evaporation steps $k$ tends to infinity.

We can prove that the overall mapping from the initial state of the black hole mass to the final state of the Hawking radiation modes is effectively an isometry. 
Since only radiation will be available outside, one can consider the map from the black hole state to only the radiation state, excluding the tiny leftover black hole. 
Using the above fact in Eq.\eqref{exponential}, we can prove that the effective evolution mapping the initial state of the black hole mass to the final state of the Hawking radiation modes is, to an arbitrarily good approximation, an isometry, whereby the information about the value of the black hole mass is spread over a large number of radiation modes (see Methods for the details.)  Interestingly, this isometric encoding of the black hole mass applies also to initial states with quantum superpositions, $|\psi\>_{\rm bh}  =    \sum_{m=0}^M   g(m)\,  |m\>_{\rm bh}$, for which the information encoded in the complex amplitudes $\{g(m)\}$ is asymptotically perfectly transferred to the Hawking radiation modes, as shown in Appendix C.

\noindent{\bf Extension to black hole with subsystems}\\
We now discuss the extension of the quantum evaporation model to the scenario where the black hole has two subsystems and a nontrivial quantum superposition of its internal mass over those subsystems. The generalization therefrom to an arbitrary number of subsystems is straightforward.  In this extended model, as we will show, the initial quantum state inside the black hole--the information about the initial internal mass superposition--can in principle be retrieved from the state of the Hawking radiation modes after a sufficiently large number of radiation emissions. 
The two subsystems are labelled as A and B, and the initial state of the  black hole's matter is 
\begin{equation}
\label{eq:BHp}
    |\psi\>_{\rm bh} =\sum_ {m=0} ^{M} f(m) \ket{m}_{\rm A} \otimes \ket{M- m}_{\rm B }
     \, ,   
\end{equation}
where $m$ is the mass in subsystem A  and $f(m)$ is the corresponding complex amplitude. We shall argue that the amplitudes $\{f(m)\}$ will be transferred to the outgoing radiation.

As in the single subsystem model, we construct the evaporation process from the combination of a Hawking production and a subsequent annihilation of the negative energy modes. Pair production takes place in the same way as before: at the first evaporation step, the Hawking pair is produced in the state $|\Phi_M\>$ in Eq. (\ref{eq:Main2}).  Note that this state depends only on the initial total mass of the black hole, $M=m_{\rm A}+m_{\rm B}$.

We assume that with every radiation burst, the infalling mode of the radiation has an amplitude to annihilate either part of the black hole mass, A or B. The annihilation of mass inside the black hole is thus described by the linear operator $W$ that transforms the state $|m_{\rm A}\>_{\rm A}  |M-m_{\rm A}\>_{\rm B}  |-\omega_1,-\omega_2\>_{\rm int}\ket{\omega_1,\omega_2}_{\rm out}$ into the state $\sum_{\omega_1=0}^{m_{\rm A}} \sum_{\omega_2=0}^{M-m_{\rm A}} q{(\omega_1,\omega_2)} |m_{\rm A}-\omega_1 \>_{\rm A}   |M-m_{\rm A}-\omega_2\>_{\rm B} \ket{\omega_1,\omega_2}_{\rm out}$. 
Note that $\omega_1$, $\omega_2$ can be one particle each or can be many lumped together.
The overall evaporation step operator is described by an operator $V_{\rm ev}$ and reads, 
\be V_{\rm ev} |\psi\>_{\rm bh} \hspace{-4mm} &&=\hspace{-2mm} \sum_{m=0}^M \sum_{\omega_1=0}^{m}  \hspace{-1mm} \sum_{\omega_2=0}^{m-\omega_1} \sqrt{p(\omega_1+\omega_2)} q{(\omega_1,\omega_2)} f(m)  \nonumber \\
&&  \ket{m - \omega_1}_{\rm A}  \ket{M- m - \omega_2}_{\rm B } |\omega_1,\omega_2\> _{\rm out}
     \, 
,
\ee
where $q$ includes the amplitudes for annihilating A or B, whereas $\sqrt{p(\omega_1+\omega_2)}$ is the standard amplitude for creating an $\omega_1 + \omega_2$ energy burst. 
The same operators are applied at later steps, but with the total mass of the black hole $m\le M$.
  
Now, suppose that $n$ evaporation steps affected A and B.   For large $n$, similar methods to those shown earlier in this paper imply that the information about the masses is nearly perfectly encoded into the Hawking radiation modes.  Specifically, in the Methods we show that the state of the radiation after a large $n$ evaporation steps is approximately of the form 
\be 
\label{eq:finalrad}
|\psi_{\rm rad} \>^{(n)}  = \sum_{m=0}^M  f^{^{(n)}}  (m)  |\Lambda_m\>_{\rm rad_{\rm A}} \otimes |\Lambda_{M-m}\>_{\rm rad_{\rm B}} ,
\ee  where $\{|\Lambda_m\>_{\rm rad_{\rm A}}\}_{m}$   ($ \{|\Lambda_{M-m}\>_{\rm rad_{\rm B}}\}_{m}$) are orthogonal states of the radiation modes that annihilated A (B).  This observation shows that the information encoded in the amplitudes $\{f^{(n)} (m) \}$ in the initial state as in Eq.~(\ref{eq:BHp}) can in principle be retrieved, with vanishing error in the limit of large $n$, from the state of the radiation.

Notice that in the case where A and B were initially entangled, the subsystems $\{|\Lambda_m\>_{\rm rad_{\rm A}}\}_{m}$ and $ \{|\Lambda_{M-m}\>_{\rm rad_{\rm B}}\}_{m}$ are entangled as well. That can be seen by comparing Eq.(\ref{eq:BHp}) with Eq.(\ref{eq:finalrad}). 

\begin{figure}
    \centering
    \includegraphics[width=\linewidth]{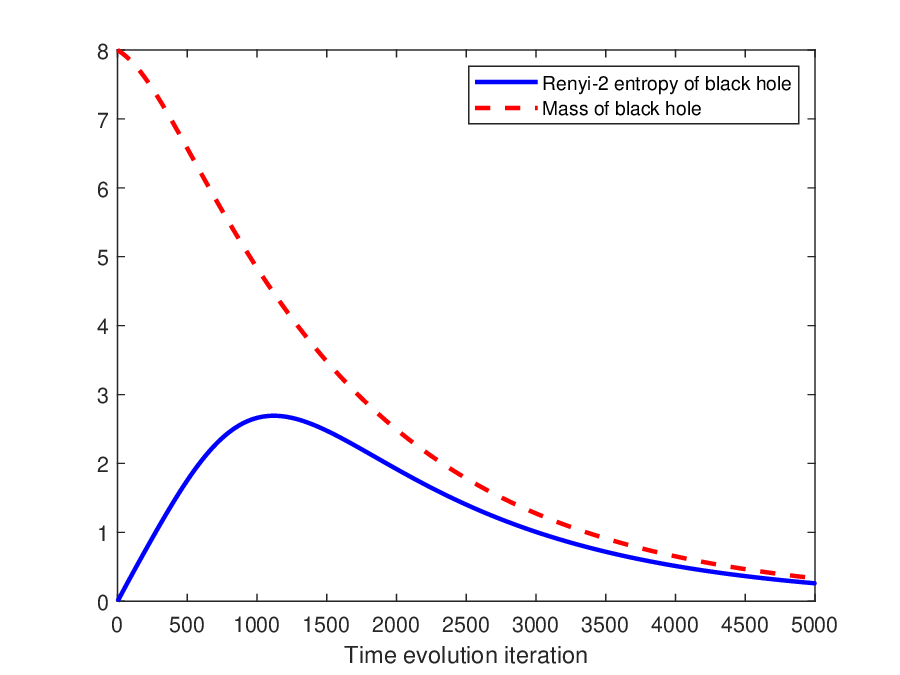}
    \caption{ {\bf Black hole evaporates and entropy of radiation vanishes.} Numerical simulation of a two-mode black hole toy model with initial state 
    $\ket{\rm bhm} =\ket{m}_{\rm A}\ket{m}_{\rm B}=\ket{4}_{\rm A }\ket{4}_{\rm B}$. The expectation value of $m_{\rm A}+m_{\rm B}$ is shown to decay, and the entropy of the black hole is shown to rise and then fall. Since the total state is pure, the entropy of the black hole equals that of the outside radiation. Simulation details are provided in the Supplementary Materials. }
    \label{fig:BHMassEntropy440} 
\end{figure}

\noindent \textbf{The Page Curve}\\
The entropy of the radiation rises and falls similarly to the Page Curve of Ref.~\cite{Page}. Fig. \ref{fig:BHMassEntropy440} shows the Renyi-2 Entropy $-\log \mathrm{tr}(\rho^2)$ for the radiation numerically computed for a specific initial state. Initially, the radiation is, by assumption, in a pure vacuum state $\ket{0}_{\rm{rad}}$ and thus has no entropy. Gradually, the radiation entropy rises as radiation is produced and entanglement forms with the black hole mass. The black hole gradually tends to a vacuum state $\ket{0}_{\rm bh}$, which implies that the entanglement with the radiation must eventually disappear. This gradual disappearance can be interpreted as entanglement between the inside and outside being swapped to solely outside entanglement, conditional on annihilation inside the black hole~\cite{Ours}. More specifically, after $k$ radiation emissions, for any $k$, the joint state of the black hole and radiation can be expressed as 
\begin{eqnarray}
\! \! \! |\psi^{(k)}_{\rm Tot}  \> \! \! &=& \! f_{\rm ev} |0\> _{\rm bh} |\Lambda^{(k)}_M\>_{\rm rad} \! \nonumber \\
&+&\!\!\! \sum_{m \neq 0} \!  f_{\rm nonev}  |m \>_{\rm bh } |\Lambda_{M-m} ^{(k)}\>_{\rm rad},
\end{eqnarray}
where $f_{\rm ev}$ is the amplitude for evaporation and $f_{\rm nonev}$ is the amplitude for non-evaporation after $k$ radiation emissions. 
At the early stage of the evaporation, newly produced pairs increase the entropy, contributing significantly to the second term of the RHS, since the probability of full annihilation $p_{\rm ev}=|f_{\rm ev}|^2$ of matter particles is very small.  On the other hand, at a much later stage of the evaporation, 
the probability of full annihilation is significantly higher and, thus, with every pair that is then created, the contribution from annihilation and entanglement transfer is larger, which decreases the entropy. The asymptotic final state has the form $\ket{0}_{\rm bh}\ket{\psi}_{\rm rad}$, where $\ket{\psi}_{\rm rad}$ is given in Eq.\eqref{eq:finalrad}. The overall behaviour is accordingly that the entropy of the radiation rises from 0, keeps rising for a significant time, and then tends later back to 0, as in Fig.~\ref{fig:BHMassEntropy440}. A rise and fall in subsystem entropy also occurs with states picked from the uniform distribution over quantum states used by Page~\cite{Page}. Three distinctions: (i) here we are not picking states or dynamics at random from the uniform distribution; (ii) the x-axis in our curve is time or number of emissions rather than the number of subsystems; (iii) subsystems are not `leaking' from the black hole in our approach, only information. 

\section{Discussion}
We presented a toy quantum model for black hole evaporation.

A key feature of our model is the {\em controlled} squeezing. The rate of squeezing is controlled by the total black hole mass. The controlled squeezing is implemented by turning the squeezing parameter into an operator, which accounts for radiation emission from a black hole with quantum coherence, allowing every branch of the black hole state superposition to emit radiation differently if it corresponds to a different total black hole mass. As every burst of radiation decreases the black hole mass, the surface gravity and the amount of squeezing at the event horizon keep updating with every emission. This controlled squeezing thus accounts for a back-reaction on the black hole and the next emitted radiation states. This back-reaction has, to our knowledge, not been previously accounted for. 

Moreover, what is perhaps the most important difference from~\cite{Hawking74,Hawking76} is that Hawking models the final state of the radiation as the reduced density matrix of the two-mode squeezed state, untouched as the black hole fully evaporates. Here, instead, we allow for a superposition of different degrees of evaporation, as for example in Eq.(\ref{eq:twostepevap}). The black hole is fully evaporated in the branches of its superposition where all of its matter particles are fully annihilated. More specifically, from two iterations of Eq.(\ref{eq:Main2}), tracing out the inside yields the thermal state
\be 
\rho_{\rm out} =\sum_{\omega_1,\omega_2}\gamma(\omega_1,\omega_2) |\omega_1\> \< \omega_1| \otimes |\omega_2 \> \< \omega_2|,
\ee
where $\gamma(\omega_1,\omega_2)=\frac{\exp(-\beta_1 \omega_1)\exp(-\beta_2 \omega_2)}{Z(\beta_1)Z(\beta_2)}$. So long as the ``interior'' of the black hole exists, it keeps the full state pure, while the outer radiation is mixed. However, the standard paradoxical conclusion that normally follows is that when the inside is fully evaporated, the Hawking radiation state is left in a mixed state with the black hole no longer existing to purify it.
In our approach, instead, we expand the black hole and radiation state in Eq.(\ref{eq:twostepevap}) into two branches, an evaporated part and a non-evaporated part. 
Conditioning on the branch where the black hole has fully evaporated, no horizon exists, and the inside and outside are disentangled. Tracing out the inside conditional on that case yields a pure state outside as it is straightforward to see. On the other hand, in the branches where the evaporation is not completed, more pairs will be created, until there is full evaporation inside, and a pure state outside, in every branch. Eq.(\ref{exponential}) makes it clear that after many bursts of radiation, this happens with a probability approaching 1.

In order to further compare our findings to previous literature, we now analyze the issue of the monogamy of entanglement. Monogamy implies that the `out' Hawking particles cannot be entangled with each other to purify the radiation state, because each of those particles is almost maximally entangled with its negative energy partner inside the black hole, or to the black hole matter that it interacted with \cite{Mathur, AMPS}. This is circumvented here by the fact that the entanglement existing within the black hole matter particles is only fully transferred to the `out' radiation upon the full annihilation of the matter particles carrying it. This mechanism does not violate monogamy, for the simple reason that at any given moment, the evaporation process is described by a quantum state that respects monogamy relations. More precisely, the global state is a superposition in which, in some branches, the early and late Hawking radiation are entangled with the black hole interior, forming a multipartite entangled state. In other branches, the entanglement has been fully swapped from the interior to the exterior and is shared between the early and late radiation modes. Crucially, these two possibilities do not coexist within any single branch — they occur in distinct branches of the superposition. This is evident comparing the two branches of the superposition in Eq.(\ref{eq:bothexplicit}). \\ 

Let us further consider the case of firewalls \cite{AMPS,Braunstein}.  
The argument therein, is that given a virtual surface, if there is no entanglement across the surface, its Hamiltonian diverges implying that its energy gets extremely high \cite{Braunstein}, whence the firewall. Our model avoids the firewall by eliminating the need to break the entanglement across the horizon, as the entanglement is swapped only upon annihilation of each internal subsystem.

A natural question is whether the model violates causality restrictions associated with the black hole event horizon. One restriction is that black hole matter cannot exit the horizon. That restriction is respected here, as the dynamics involve no matter escaping the black hole. Another type of restriction concerns information transfer. We first distinguish here between total mass and the relative mass superposition information. The black hole's total mass is accessible from the outside due to the black hole's gravitational field, in line with the no-hair theorem ~\cite{Landau}. The relative mass superposition information, however, needs an elaborate discussion.

Information gradually escapes throughout the evaporation process. In the intermediate stage, the global state is a superposition of branches: some in which the black hole has fully evaporated and the initial information has escaped into the external radiation, and other branches in which the black hole has not yet evaporated, in which case the branch state is still an entangled state between the interior and radiation modes. If, at any point during the evolution, an external observer measures the radiation, they would be able to recover a partial amount of information about the black hole matter state. This may or may not be tantamount to signalling, a question that needs a more 'spacetime-elaborate' model. However, in the current state of our model, tracing out the black hole interior degrees of freedom yields a reduced density matrix for the radiation that depends on the detailed quantum state of the black hole matter.  In particular, both the fully evaporated branch—where all information has escaped—and the partially evaporated branches—where only some radiation has been emitted—carry information about the black hole’s initial state. Note that this does not necessarily contradict the standard semiclassical result that no information can escape in the presence of a fixed background with a horizon. In our case, the background is not fixed but exists in a quantum superposition. Moreover, once a measurement is performed on the emitted radiation, the global quantum state generally changes. In the extreme case, if one were to perform frequent measurements on a branch in which the black hole has not evaporated, the evolution could be effectively frozen (quantum Zeno effect), resulting in a black hole with a fixed horizon from which no information escapes thereafter. Hence, in our model, no information escapes until radiation is actually emitted. This implies that the only mechanism for information to leave the black hole is through radiation emission. The process continues until the black hole fully evaporates, at which point all the information can, in principle, be recovered outside.

\section{Methods}

\noindent \textbf{Hawking radiation and squeezing}\\
We model the evaporation as a sequence of Hawking radiation creations rather than a single squeezing operation. Hawking compared the vacuum states' definitions in the black hole background geometry at the future null-like infinity $\mathcal J^+$ with the Minkowski flat spacetime geometry at the past null-like infinity $\mathcal J^-$. The radiation state thus derived by Hawking and others is a two-mode squeezed state \cite{Hawking74, Hawking76, HawkingInfo}:
\begin{equation}
\label{eq:MainMeth}
    \ket{\rm HR} = \bigotimes_{\omega > 0} c_\omega \sum_{N_\omega=0}^{\infty} {\rm e}^{- \pi N_\omega \omega M } \ket{ N_{-\omega}}_{\rm int}  \otimes \ket{ N_\omega}_{\rm out} \, , 
\end{equation}
where $c_\omega \equiv \sqrt{1- {\rm e}^{- 2 \pi \omega M }}$ is a normalization factor, 
$N_\omega$ is the number of particle pairs of energy $\omega$, while ``int'' and ``out'' label the Hilbert spaces for the particles falling inside the black hole and those escaping to the future infinity, respectively~\cite{Fabbri}.

The state of Eq.\eqref{eq:MainMeth} is expressed in the number eigenbasis. Keeping track of all the numbers over all the energies is possible, but will unnecessarily complicate our notation without affecting our results. For the sake of simplicity, we will focus on a  single excitation pair with indefinite energy $\omega$, which is straightforward to derive from Eq.\eqref{eq:MainMeth} and reads,
\begin{equation}
\label{eq:1pair}
    \ket{\rm \phi_M} = \sqrt{1 - {\rm e}^{-  2 \pi M}} \sum_\omega ^{\infty} {\rm e}^{-  \pi M \omega }  \ket{-\omega}_{\rm int} \otimes  \ket{ \omega} _{\rm out} ,  
\end{equation}
where $\sqrt{1 - {\rm e}^{-  2 \pi M}} $ is a normalization factor.  

Our model incorporates repeated bursts of Hawking radiation. The Hawking state of Eq.(\ref{eq:MainMeth}) is commonly thought of as the full state of all radiation throughout the evaporation process. However, by a simple expansion of the tensor product and the sums one can see that it is insufficient for full evaporation. The terms in the expansion with the highest probabilities have the form: $|1_{\omega_1}, 0_{\omega_2},0_{\omega_3},... \> , ~ |0_{\omega_1}, 1_{\omega_2},0_{\omega_3},... \> , ~ {\rm etc...} $ with the lowest energies $\omega_j$ being the most occupied. Thus, in the branches of the superposition with the highest probabilities, the black hole has its mass reduced the least. The radiation emission process is therefore expected to continue. 
We will treat each emission of Hawking radiation (more specifically, each creation of an energy-truncated version of Eq.\eqref{eq:MainMeth} or, equivalently, Eq.\eqref{eq:1pair}) as a single burst of radiation, with the full evaporation process constituting multiple such bursts.

Both equations (\ref{eq:MainMeth}) and (\ref{eq:1pair}) are the results of applying the squeezing operator on the vacuum state. The squeezing operator reads, 
\begin{equation}
\label{eq:squeeze1}
    \mathcal{ S}(k,M)= \ex^{\zeta(k,M) \left( a^{\rm int\dagger}_{- k} a^{\rm out \dagger} _k  - a^{\rm int}_{- k} a^{\rm out}_k \right) } \, ,
\end{equation}
where $\zeta(k,M) =  \arctan{ \left( \ex^{- \pi k M} \right)}$, $k$ is the energy of the emitted particles, and $M$ is the total black hole mass. $M$ is a scalar for now.
$\mathcal S$ acts on the `int' and `out' radiation Hilbert spaces, creating a negative energy flux in the interior, and a positive energy flux outside.  
 Now, recall that the operator $\mathcal S$ acts on the vacuum state of the radiation mode. In this subspace, its action is equivalent to a unitary operator 
\begin{equation}
\label{eq:squeeze-eps}
    \begin{aligned}
         \mathcal S(k,M) & =   \frac{1}{\cosh \zeta(k,M)} \exp( (\tanh\zeta) a^{\rm int\dagger}_{- k} a^{\rm out \dagger} _k ) \\
         = & \sqrt{1 - {\rm e}^{-2 \pi k M} }\sum_{n=0}^\infty  \frac{{\rm{e}}^{- \pi  n k M }}{n!}  \left(a^{\rm int\dagger}_{- k} a^{\rm out \dagger} _k \right) ^{n}
    \end{aligned}
\end{equation}
(see Supplementary Note~\ref{sec:squeezed} for the derivation). 
Instead of restricting the state in Eq.\eqref{eq:MainMeth} to one pair states, an alternative way to get Eq.\eqref{eq:1pair} is applying the squeezing operator for the quantum of energy $m_*$ to the vacuum state, and interpreting $n m_* \equiv  \omega$ as the energy of the produced pair. That is,  
\be \hspace{-5mm} S(m_*,\! M)|0\> \hspace{-1mm}  & = \hspace{-1mm} \sum_{n=0}^\infty \hspace{-1mm} \sqrt{p_M(n m_*)} |\hspace{-0.5mm}  -\!n m_*  \>_{\rm int} |n m_* \>_{\rm out}~. \label{eq:quanta}
\ee 
In this case, there are $n$ excitations of energy $m_*$ each; these $n$ excitations can be one or more particles. In the case of a single system black hole, we will treat all the $n$ excitations as one system, which will be sufficient for our purposes. That is, $S(m_*,M)\ket{0}_{\rm int, out}=\ket{\phi_M}$. In the two subsystem black hole case, we label the quanta that annihilate A as $n_1 m_* \equiv \omega_1$, and the $n_2= n- n_1$ quanta that annihilate B as $(n- n_1) m_* \equiv \omega_2 \equiv \omega - \omega_1$. 

The product of all possible energies squeezed from the vacuum $\displaystyle \otimes_{k} \mathcal S(k,M)$ gives the number eigen-basis Hawking state as in Eq.~\eqref{eq:MainMeth}, as presented in \cite{Fabbri}.
We will generate the state for $n$ quanta $m_*$ at each elementary time step. The state is essentially that of (Eq.(\ref{eq:1pair})), though we will truncate the sum for radiation energies that are higher than the black hole mass, assuming that the black hole cannot create more radiation than what it has, which would violate energy conservation. 
The radiation state we are left with will therefore be Eq.(\ref{eq:Main2}), which has a different summation limit.

\noindent \textbf{Controlled squeezing}\\
To enact squeezing in the context of a quantum superposition of black hole masses, we introduce a controlled squeezing operator. The Hawking radiation state depends on the total black hole mass $m$ (see Eq.~\eqref{eq:MainMeth}). However, this mass varies with every burst of radiation (see Eq.~\eqref{evaporationstep}) as the black hole loses an amount of energy equal to that of the radiation it emits. In the original computations by Hawking, this back-reaction is not accounted for. Here we make use of controlled unitaries (in the Quantum Circuits sense \cite{NielsenChuang}), to account for this back-reaction. With every burst of radiation, the black hole loses an indefinite amount of mass, and the next burst will be created with a squeezing parameter that depends on the new mass. The particle production is therefore an operation that is controlled over the black hole's total mass, or more generally, the inverse of the surface gravity evaluated at the event horizon, which equals the inverse of the black hole mass only for the case of a Schwarzschild black hole. That is,
\begin{equation}
    \kappa_{\rm Sch} = \frac{1}{M_{\rm Sch} } \,.
\end{equation}
Moreover, as we discussed above, the black hole evolves into a joint superposition with the Hawking radiation and, in principle, does not have a definite mass. Therefore, the squeezing parameter should be different for different branches of the black hole and radiation superposition. This issue is resolved by the quantum-controlled operation. The controlled operation allows the different branches of the black hole mass superposition to emit radiation with different squeezing parameters, thereby adding an extra component of quantumness to the evaporation process.

The controlled-squeezing operation reads
\begin{equation}
\label{eq:unitary-pair-creation}
    \begin{aligned}
       \mathcal{C_{\rm S }} & =   |0\>\<0|_{\rm bh} \otimes \tI_{\rm int , \: out}  \\ 
       & + \sum_{m > 0} |m\>\<m|_{\rm bh} \otimes \mathcal S{(m)}_{\rm int \: out} \, ,
    \end{aligned}
\end{equation}
where
\begin{equation} 
    \mathcal S(m)_{\rm int \; out} = \ex^{\zeta(m) \left( a^{\rm int \dagger} b^{\rm out \dagger}   - a^{\rm int} b^{\rm out} \right) } \, ,
\end{equation}
with $\zeta(m)\equiv \zeta(m_*,m)$ defined after Eq.(\ref{eq:squeeze1}). $\mathcal I$ is the identity operator and $\mathcal{C_{\rm S }}$ acts on the black hole matter and its total mass, then applies mass-dependent exponentially suppressed particle production. 
As we sum over all values of that total mass in the operator, $\mathcal C_{\mathcal S}$ is manifestly unitary.
We assume, however, that if it happens that the radiation energy inside and outside exceeds the black hole mass (which can only happen with the final radiation burst), then the positive and negative radiation left-overs will cancel each other.
Another (more general) way to write the same unitary is to let the control system be the black hole horizon, as its surface gravity is exactly the squeezing parameter. 
Hence,
\begin{equation}
\label{eq:unitary-pair-creation2}
    \begin{aligned}
     \mathcal{C_{\rm S }}  =& \, |\kappa = 0 \>\< \kappa =0 |_{\rm bh} \otimes \tI_{\rm int , \: out}  \\ 
    &+ \sum_{\kappa > 0} |\kappa \> \< \kappa |_{\rm bh} \otimes \mathcal S(\kappa)_{\rm int \: out} \, ,
\end{aligned}
\end{equation}
where $\mathcal S(\kappa)_{\rm int \: out}$ is the same squeezing operator.

We shall implement the quantum controlled squeezing via an associated isometry $V_{\rm prod}$, where the out modes are by choice of convention initially not part of the input state. By definition $V_{\rm prod}\ket{\psi}_{\rm bh} \equiv \mathcal C_{\mathcal S} \left( \ket{0}^{\rm int} \ket{0}^{\rm  out} \ket{\psi}_{\rm bh}\right) $. As an example to illustrate how the controlled squeezing then works, below we let $V_{\rm prod}$ act on a black hole that is in a superposition of being of energies 0, $m_1$, and $m_2$.  Namely, we have
\be
&& V_{\rm prod} \frac{1}{\sqrt 3}(|0\>_{\rm bh} + |m_1\>_{\rm bh} + |m_2\>_{\rm bh} ) \nonumber \\
&& =  \frac{1}{\sqrt 3}\bigg(|0\>_{\rm bh}|0\rangle_{\rm int} |0\rangle_{\rm out} +\nonumber\\  && \sum_{k_1} \sqrt{p_{m_1}(k_1) } |m_1\>_{\rm bh}|-k_1\>_{\rm int} |k_1\>_{\rm out} +\nonumber \\
&& \sum_{k_2} \sqrt{p_{m_2}(k_2)}  |m_2\>_{\rm bh} |-k_2 \> _{\rm int} |k_2\>_{\rm out} \bigg) .
\ee
Notice that for the branches of the superposition where the black hole has fully evaporated, there is no longer any particle production, whereas in the branches where the black hole has not fully evaporated, there are new particles that are produced and others annihilated.
The probabilities for full or partial annihilation are thus updated sequentially.

\noindent {\bf  Combined operator for radiation production and annihilation with black hole matter}  \\
In the quantum evaporation model, the evaporation process arises from two intermediate steps:  the production of Hawking pairs and the annihilation of the black hole matter. These two intermediate steps involve quantum particles with negative energy inside the black hole and we regard them as effective processes, useful to clarify the logic of our model, but not interpreted as fundamental physical processes. 

The production of a Hawking pair is described by a linear operator  $V_{\rm prod} :  \spc H_{\rm bh}  \to   \spc H_{\rm bh} \otimes   \spc H_{\rm in} \otimes \spc H_{\rm out}$, where   $\spc H_{\rm bh}$,  $\spc H_{\rm in} $, and $\spc H_{\rm out}$ are the Hilbert spaces of the black hole matter, incoming mode, and outgoing mode, respectively.  The action of the production operator is uniquely determined by the relation  
\begin{equation}
\label{eq:Vprod}
V_{\rm prod}|m\>_{\rm bh}  = |m\>_{\rm bh} |\Phi_m\>_{\rm in, out},
\end{equation}
with $|\Phi_m\>_{\rm in , out}  = \sum_{\omega =0}^m  \,  \sqrt{p_m (\omega)}  |-\omega\>_{\rm in}   |\omega\>_{\rm out} $, 
\begin{align}\label{pmomega}
p_m(\omega)  := |c_m|^{2} {\rm e}^{-2\pi  m \omega} \,,
\end{align}
and $|c_m|^2  = \frac{1}{ \sum_{\omega=0}^m  {\rm e}^{-2\pi  m \omega}} $. $V_{\rm prod}$ equates to an energy-truncated version of the controlled squeezing of Eq.\eqref{eq:unitary-pair-creation}.  From the definition of Eq.~\eqref{eq:Vprod}, it is immediate to see that the production operator $V_{\rm prod}$ is an isometry, that is, it satisfies the condition  $\<\psi | V_{\rm prod}^\dag V_{\rm prod} |\psi'\> =  \<  \psi| \psi'\>$ for every pair of vectors $|\psi\>,  |\psi'\> \in  \spc H_{\rm bh}$.  This property guarantees that $V_{\rm ev}$ is invertible and preserves the information about the initial state of the black hole mass.

\begin{figure}[h]
\centering
\vspace{1cm}
\[
\Qcircuit @C=2.7em @R=2.2em {
    \lstick{\text{A}} & \multigate{2}{\mathcal{G}} & \ctrl{1} & \multigate{1}{W} & \qw \\
     & \nghost{\mathcal{G}} & \multigate{1}{\mathcal{S}}_{\text{int}} & \ghost{W} \\
     & \nghost{\mathcal{G}} & \ghost{\mathcal{S}} & \qw
     & \rstick{\rm{out}}\qw
}
\]
\vspace{0.10cm}
\begin{center}
$=$
\end{center}
\vspace{0.15cm}
%
\begin{tikzpicture}
  \draw (0.5,0.3) -- (2,0.3) node[midway,above] {$\text{A}$};
  \draw (2,-0.5) rectangle (3,0.5) node[midway] {$V_{\text{ev}}$};
  \draw (3,0.3) -- (4,0.3) node[right] {$\text{A}$};
  \draw (3,-0.3) -- (4,-0.3) node[right] {$\text{out}$};
\end{tikzpicture}

\vspace{0.5cm}
\caption{{\bf The combined evaporation operation.} The circuit diagram shows the equivalence between a sequence of operations (creation of auxiliary int systems via $\mathcal{G}$, controlled-$S$ operation, and partial projection $W$) and a single map $V_{\rm{ev}}$ on system A and out. In this simple case, there is a single system A representing the black hole. The Hawking pair particles are $int$ and $out$.}
\label{fig:circuit-equivalence}
\end{figure}
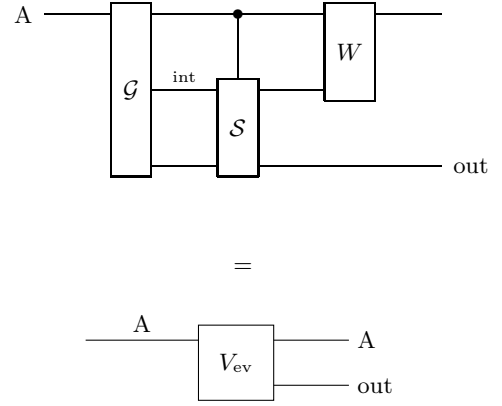

The annihilation of black hole matter is described by the linear operator $W:  \spc H_{\rm bh} \otimes \spc H_{\rm int}  \otimes \spc H_{\rm out} \to  \spc H_{\rm bh} \otimes \spc H_{\rm out}$ uniquely specified by the relation  $W  |m\>_{\rm bh} |\omega\>_{\rm int} |\omega\>_{\rm out}  =  |m-\omega\>_{\rm bh} |\omega\>_{\rm out}, \forall \omega \le m.$    
 This relation shows that the annihilation operator is not an isometry.  However, the combination of the production and annihilation operators, depicted in Fig.\ref{fig:circuit-equivalence}, yields an isometry 
\be \label{eq:Vev}
V_{\rm ev}  :  =  (W\otimes I_{\rm out})   V_{\rm prod},
\ee
 where $I_{\rm out}$ denotes the identity operator on $\spc H_{\rm out}$. The fact that $V_{\rm ev}$ is an isometry can be seen from the explicit expression 
 \begin{align}\label{Vev1} 
 V_{\rm ev}  =  \sum_m  |\Psi_m\>_{\rm bh, out}  \< m|_{\rm bh} \,,
 \end{align}
 with 
\begin{align}\label{psim}|\Psi_m\>   =    \, \sum_\omega  \sqrt{p_m  (\omega)}  \,  |m-\omega\>_{\rm bh}  |\omega\>_{\rm out} \,,
\end{align}
as $V_{\rm ev}$ sends an orthonormal basis to an orthonormal basis. 

It is important to note that we model the annihilation with the linear map $W$ in a way that captures the intuition that the energy of the emitted radiation is taken out from the black hole matter energy. Although this is a natural assumption, more in-depth work is required to find out how exactly the annihilation happens. One potential mechanism, established by Sorkin in Ref.~\cite{Sorkin} and further developed by Adami in Ref.\cite{Adami,bhinfobook}, involves a beam splitting operator replacing or supplementing the squeezing operator, replacing the usual creation of negative energy particles with annihilation of positive energy particles. That is achieved through a unitary operator acting on the infalling black hole matter and the outgoing radiation. Another possible route for understanding the detailed dynamics can come from the fact that the `int' particle, having a negative energy, potentially makes it a ghost particle (very common in the Standard Model of particle physics \cite{Weinberg}) that is off-shell (not directly observable) for an observer outside the horizon. This needs to be explored further with all its technicalities. In any case, the production and annihilation steps in our model are not to be taken as sequential elementary operations. Rather, $V_{\rm ev}$ is treated as the elementary operation.  Appendix \ref{app:Signalling} shows that if $V_{\rm ev}$ is treated as a sequence of two fundamental operations $V_{\rm prod}$ and $W$, that would lead to the possibility of signalling.

\noindent \textbf{The quantum state after $k$ steps}\\ 
A sequence of $k$ evaporation steps corresponds to $k$ subsequent applications of the evaporation operator, and the overall evolution is given by the operator $V_{\rm ev}^{(k)}  :  \spc H_{\rm bh}  \to  \spc H_{\rm bh}  \otimes  \spc H_{\rm out_1} \otimes  \spc H_{\rm out_2} \otimes  \cdots \spc H_{\rm out_k}$ defined by $V_{\rm ev}^{(k)}  := (V_{\rm ev} \otimes I_{\rm out_1}  \otimes \cdots \otimes I_{\rm out_{k-1}})  \cdots  ( V_{\rm ev}  \otimes I_{\rm out_1} )  V_{\rm ev}.$    

After $k$ evaporation steps, the black hole mass is entangled with the energy of  $k$ radiation modes, and their  joint state can be written as 
\begin{align}\label{psik}
|\Psi_M^{(k)}\>_{{\rm bh, rad} _k}    = 
\sum_{m=0}^M  \,  f^{(k)}_M (m) ~ \,  |m\>_{\rm bh}  |\Lambda^{(k)}_{M-m}\>_{\rm rad_k}  \,,
\end{align}
where $f^{(k)}_M (m)$ are suitable amplitudes, $\rm rad_k : = out_1 \cdots out_k$ is a shorthand notation for the $k$ radiation modes, and  $|\Lambda^{(k)}_{M-m}\>_{\rm rad_k}$ is a suitable state with total energy $M-m$. 

At the $k$-th  step, the probability of complete evaporation is 
\begin{align}\label{pevm}
\nonumber p_{\rm ev}^{(k)}  (M)  &=\<  \Psi^{(k)}_M  |  \left(|0\>\<0|_{\rm bh}  \otimes I_{\rm rad_k}\right)  |\Psi^{(k)}_M\>  \\
&=    \left|  f_M^{(k)}  (0)\right|^2 \,.
\end{align}
In the following, we show that this probability tends to 1 in the large $k$ limit.  For this purpose,  we establish a recurrence relation for $p_{\rm ev}^{(k)}  (M)$.   Using Eq. (\ref{psik}), the state after the $(k+1)$-th evaporation step can be written as 
\begin{align}
\nonumber |\Psi_M^{(k+1)}\>_{{\rm bh, rad}_k} &   = 
\sum_{m=0}^M  \,  f^{(k)}_M (m)  \,  \left( V_{\rm ev}|m\>_{\rm bh} \right)  \,  |\Lambda^{(k)}_{M-m}\>_{{\rm rad}_k} \\
 &\quad= 
\sum_{m=0}^M \sum_{\omega= 0} ^{m} f^{(k)}_M (m)  \,\sqrt{p_m  (\omega)}  \,|m-\omega\>_{\rm bh} \nonumber    \\
&\hspace{1.2cm} \otimes |\Lambda^{(k)}_{M-m}\>_{{\rm rad}_k}  |\omega\>_{{\rm out}_{k+1}} \,,
\label{psik+1}
\end{align}
where the last step follows from Eqs. (\ref{Vev1}) and (\ref{psim}).  
Hence, the probability of full evaporation at the $(k+1)$-th step is 
\begin{align}
\nonumber p_{\rm ev}^{(k+1)}  (M) &  =  \sum_{m=0}^M  \,  |f^{(k)}_M  (m)|^2 \, p_m   (m) \\
&=    p_{\rm ev}^{(k)}  (M)  +  (1-p_{\rm ev}^{(k)}  (M)) ~ \<   p_m (m)\>    \, , \label{bound1}
\end{align}
with $\<   p_m (m)\>   :  = \sum_{m  =  m_*}^M   p_m (m)  |f^{(k)}_M  (m)|^2/(1-p_{\rm ev}^{(k)}  (M)).$  Now, Eq. (\ref{pmomega}) implies the expression   $p_m (m)  =  {\rm e}^{-2 \pi m^2}/(  \sum_{\omega=0}^m   {\rm e}^{-2\pi  m\omega}) $, which implies  $p_m(m)\ge  p_M (M)$ for every $m\le M$. This relation implies the relation $\<  p_m (m)\>  \ge p_M(M)  \equiv p_{\rm ev}^{(1)}  (M)$ for the expectation value of $p_m(m)$.   Inserting this relation into Eq. (\ref{bound1}), we then obtain  the bound
\begin{align}
p_{\rm ev}^{(k+1)}  (M) \le     p_{\rm ev}^{(k)}  (M)  +  (1-p_{\rm ev}^{(k)}  (M))    \, p_{\rm ev}^{(1)}  (M)     \, , 
\end{align}
which in turn implies the recursion relation 
\begin{align}
1- p_{\rm ev}^{(k+1)}  (M) \le  \left(1-   p_{\rm ev}^{(k)}  (M) \right) \, \left( 1-  p_{\rm ev}^{(1)}  (M) \right)    
\end{align}
and the bound 
\begin{align}\label{recursion}
1- p_{\rm ev}^{(k+1)}  (M) \le  \left( 1-  p_{\rm ev}^{(1)}  (M) \right)^{k+1} \, .    
\end{align}
To conclude, we evaluate the r.h.s. using Eq. (\ref{pmomega}), which yields
\begin{align}
\nonumber 1-  p_{\rm ev}^{(1)}  (M)    &=  \frac{\sum_{\omega = 0}^{M-  m_*}    {\rm e}^{-2\pi  M\omega}}{  \sum_{\omega'=0}^M {\rm e}^{-2\pi M \omega}}\\
\nonumber &  = \frac{      (  1  - {\rm e}^{-2\pi  M^2} )}{1-  {\rm e}^{-2\pi  M  (M+ m_*)}}\\
& \le 1  -  {\rm e}^{-2\pi  M^2}  + {\rm e}^{-2\pi  M  (M+ m_*)}  \, . 
\end{align}
Inserting this bound into the recursion relation (\ref{recursion}), we finally obtain 
 \be p_{\rm ev}^{(k+1)}  (M)  \ge 1  - \left(  1  -  {\rm e}^{-2\pi  M^2}  + {\rm e}^{-2\pi  M  (M+ m_*)} \right)^{k+1} \hspace{-2mm} ,\ee
 thereby proving Eq. (\ref{exponential}). 

Since the probability of evaporation converges to $1$ in the large $k$ limit, Eq. (\ref{pevm}) implies that the state $|\Psi_M^{(k)}\>$ has the approximate expression $|\Psi_M^{(k)}\>\approx |0\>_{\rm bh}  \otimes |\Lambda^{(k)}_M\>_{\rm rad_k}$, where  $|\Lambda^{(k)}_M\>_{\rm rad_k}$  is a suitable state of the radiation modes, with total energy $M$.    The same reasoning applies to all the states $|\Psi_m^{(k)}\>:  =  V^{(k)}_{\rm ev}  |m\>_{\rm bh}$, which can be shown to have the approximate expression  $|\Psi_m^{(k)}\>\approx |0\>_{\rm bh}  \otimes |\Lambda^{(k)}_m\>_{\rm rad_k}$.  This fact suggests that the effective evolution from the initial states of the black hole mass to the final states of the radiation modes is approximately an isometry, of the form $V_{\rm eff}^{(k)}  =  \sum_{m=0}^M  \,  |\Lambda_m^{(k)}\>_{\rm rad_k}\<m|_{\rm bh}$.  
  In Supplementary Note \ref{app:isometry}, we provide a rigorous derivation, showing that the effective channel $\map C_{\rm eff}^{(k)} (\cdot)  :  = \Tr_{\rm bh}  [  V_{\rm ev}^{(k)}  (\cdot) V_{\rm ev}^{(k)  \dag}]$,  transforming black hole states into radiation states satisfies the condition 
  \begin{align}
  \|  \map C_{\rm eff}^{(k)}    -   \map V_{\rm eff}^{(k)} \|_\diamond  \le   2 \sqrt{  2 \epsilon_k  }  \,,
  \end{align}
where $\|  \cdot \|_\diamond$ is the diamond norm \cite{kretschmann2006information}, and  $\map V_{\rm eff}^{(k)}$ is a suitable isometric channel.    The proof in \ref{app:isometry} relies on the following.   Consider the dual channel $\map D^{(k)}  (\cdot)   : = \Tr_{\rm rad_k}  [  V_{\rm ev}^{(k)}  (\cdot) V_{\rm ev}^{(k)  \dag}]$.  We show that $\| \map D^{(k)}   -  |0\>\<0|  \, \Tr    \|_\diamond \le \epsilon_k$, for some $\epsilon_k$ related to the probability of evaporation. Then, it follows from the continuity of the Stinespring dilation that the isometric channel with isometry $V_{\rm ev}^{(k)}$ is close to an isometric channel with isometry $|0\>_{\rm bh} \otimes V_{\rm eff}^{(k)}$. Finally, taking the trace over $\spc H_{\rm bh}$ shows that $\map C_{\rm eff}^{(k)}$ is close to $\map V_{\rm eff}^{(k)}$.

\noindent
\textbf{Preservation of information about the internal mass superposition}  \\
Here we consider the case of a black hole of total mass M, made of two subsystems, A and B. The most general initial black hole state can be written as,
\be 
| \psi \rangle_{\rm A B} = \sum_m f(m) |m \> _{\rm A} |M - m\>_{\rm B} \, .
\ee
The production of a Hawking pair is described by a linear operator  $V_{\rm prod}  :  \spc H_{\rm A} \otimes \spc H_{\rm B}  \to   \spc H_{\rm A} \otimes \spc H_{\rm B} \otimes   \spc H_{\rm int} \otimes \spc H_{\rm out}$, uniquely determined by the relation  $V_{\rm prod}|m\>_{\rm A} |M-m\>  = |m\>_{\rm A} |M-m \>_{\rm B} |\Phi_M\>_{\rm int, out}$, with $|\Phi_M\>_{\rm int , out}  = \sum_{n =0}^{M/m_*}  \,  \sqrt{p_M ( n m_*)}  |-n m_*\>_{\rm int}   |n m_*\>_{\rm out} $, $p_M(\omega)  := |c_M|^2 {\rm e}^{-2\pi  M \omega}$, 
and $|c_M|^2  = \frac{1}{ \sum_{\omega=0}^M  {\rm e}^{-2\pi  M \omega}} $. Although this $n m_*$ is expressed as a ket, here the number of particles that it consists of is not specified. That is because we do not keep track of particle numbers and rather focus on the energy basis. In the following analysis, we will lump the part of omega that interacts with A in a ket and the part that will interact with B in another ket. 

When annihilation takes place inside the black hole, we keep a general case of a superposition of annihilating A and annihilating B, with amplitudes $q(\omega_1,\omega_2)$. The explicit form of $q$ depends on the inner black hole state, in our case fully captured by $f(m)$. 
Then, the state $|m_{\rm A} \>_{\rm A}  |-\omega\>_{\rm int}|\omega\>_{\rm out}$ is mapped into $\sum_{\omega_1=0}^{m_{\rm A}} \sum_{\omega_2=0}^{M-m_{\rm A}} \sqrt{p_M{(\omega_1+\omega_2)}} q{(\omega_1,\omega_2)} |m_{\rm A}-\omega_1 \>_{\rm A}   |M-m_{\rm A}-\omega_2\>_{\rm B} \ket{\omega_1,\omega_2}_{\rm out}$. Where $\sqrt{p_M{(\omega_1+\omega_2)}}=  c_{M} {\rm e}^{-M (\omega_1 + \omega_2) } $. 
Overall, the above model implies that an evaporation step affecting the black hole subsystems corresponds to an application of the isometry $V_{\rm ev}:   \spc H_{\rm A} \otimes \spc H_{\rm B}  \to \spc H_{\rm A} \otimes \spc H_{\rm B} \otimes \spc H_{\rm out}$, uniquely defined by the relation  
\be 
&& \hspace{-3mm} V_{\rm ev} (|m_{\rm A}\>_{\rm A} |m-m_{\rm A}\>  _{\rm B} )\nonumber \\ 
&& \hspace{-2mm} \equiv \! |\Psi^{({\rm A},m)}_{m_{\rm A}} \> \!_{\rm A ,out}   |\Psi^{(B,m)}_{m-m_{\rm A}} \> _{\rm B ,out} \\
&&  \hspace{-2mm} =  \hspace{-1mm} \sum_{\omega_1=0}^{m_{\rm A}} \hspace{-1mm}  \sum_{\omega_2=0}^{M-m_{\rm A}} \hspace{-2mm}  \sqrt{p_M(\omega_1+\omega_2)} q{(\omega_1,\omega_2)} |m_{\rm A}\!-\!\omega_1\>_{\rm A}   \nonumber \\ 
&& \hspace{2.4cm} \otimes |m\!-\!m_{\rm A}\!-\!\omega_2\>_{\rm B}  |\omega_1,\omega_2\>_{\rm out} \label{eq:oneemission}
, 
\ee
with $\omega_1+\omega_2 = \omega$ being the emitted radiation's energy at one burst. 

It is important here to explore what the evaporated state looks like. For that purpose, we split the sum in Eq.\eqref{eq:oneemission} into the evaporated branches and non-evaporated branches. 
The state then reads,
\be \label{eq:2branches}
&& \hspace{-3mm} V_{\rm ev} (|m_{\rm A}\>_{\rm A} |m-m_{\rm A}\>  _{\rm B} )\nonumber \\ 
&&   =  \! \! \sqrt{p_m(\omega_1+\omega_2)} q{(m_{\rm A},\! m \! - \! m_{\rm A}\! )} |0\>_{\! \rm A} |0\>_{\rm \!B}  |m_{\rm A},\! m\! -\! m_{\rm A}\>_{\hspace{-0.4mm} \rm out}  \nonumber \\ 
&& + \hspace{-2mm} \sum_{\omega_1=0}^{m_{\rm A}-m*}   \, \sum_{\omega_2=0}^{M-m_{\rm A}-m*} \hspace{-4mm}  \sqrt{p_m(\omega_1+\omega_2)} q{(\omega_1,\omega_2)} |m\!-\!\omega_1\>_{\rm A} \nonumber \\ 
&& \hspace{2.4cm} \otimes |m\!-\!m_{\rm A}\!-\!\omega_2\>_{\rm B}    |\omega_1,\omega_2\>_{\rm out} ~.
\ee
As is evident in Eq.\eqref{eq:2branches}, in the branches where the black hole has fully evaporated, the quantum state outside the black hole is pure and carries information about the black hole matter state. 


Now, every application of the evaporation isometry $V_{\rm ev}$ increases the probability that the mass in subsystem A (as well as subsystem B) is fully evaporated. For example, consider the effect of one application of $V_{\rm ev}$ to a generic state 
\begin{eqnarray}
    \label{eq:GenInitial} &|\Psi\>_{{\rm A ,B ,   rad}_n}&  =\\  &\sum_{m =0 }^M  \! \sum_{m_{\rm A}  =0}^m & \!\! f(m,m_{\rm A}) \,  |m_{\rm A}\>_{\rm A}  |m-m_{\rm A}\>_{\rm B}    |\Lambda_{M-m}^{(m)}\>_{\rm rad_n},\nonumber
\end{eqnarray} 
where $f(m,m_{\rm A})$ are  arbitrary complex amplitudes, and $|\Lambda_{M-m}^{(m)}\>_{\rm rad_n}$ is a state of $n$ already emitted radiation modes, for an arbitrary $n\ge 0$.  After the new evaporation step, the state becomes    
  \begin{align} \label{eq:2ParEvap}
  \nonumber  &|\Psi'\>_{{\rm A ,B , \rm rad}_{n+1}}  :  =  V_{\rm ev}  \,  |\Psi\>_{{A ,B ,  \rm rad}_n} \\
  \nonumber & = \sum_{m =0 }^M  \sum_{m_{\rm A}  =0}^m   \sum_{\omega_1=0}^{m_{\rm A}} \sum_{\omega_2=0}^{m- m_{\rm A}} \sqrt{p_m(\omega_1+\omega_2)} q{(\omega_1,\omega_2)} f(m,m_{\rm A}) \\
  & \times |m_{\rm A}-\omega_1 \>_{\rm A}  |m-m_{\rm A}-\omega_2\>_{\rm B}    |\Lambda_{M-m}^{(m)}\>_{\rm rad_n}  |\omega_1,\omega_2\>_{\rm out_{n+1}} \nonumber 
  \, .
  \end{align}
The probability of finding subsystem A fully annihilated is then 
\begin{align}
\nonumber p_{\rm ev,A}'  & =  \sum_{m=0}^{M}  \sum_{m_{\rm A}=0}^m \sum_{\omega = m_{\rm A}}^{m} |f(m,m_{\rm A})|^2 \,  |c_{m}|^2 \,  {\rm e}^{-2\pi m \omega}\\
&=  p_{\rm ev, A}     +     \< p_{m_{\rm A}}^{(m,m_{\rm A})} \>
\,       
\end{align}
with $p_{\rm ev, A}  :  =  \sum_{m=0}^M     \,  |f(m,0)|^2$ and  $\< p_{m_{\rm A}}^{(m,m_{\rm A})} \>=  \sum_{m=m_*}^M\sum_{m_{\rm A}  =m_*}^{m}    g_{m,m_{\rm A}}   \, p_{m_{\rm A}}^{(m,m_{\rm A})}$, where $ g_{m,m_{\rm A}}$ is the probability distribution of the pair $(m,m_{\rm A})$, conditional on the fact that $m_{\rm A}$ is nonzero.

The initial preparation of the states in Eq.~\eqref{eq:GenInitial} is encoded in the amplitudes $f(m,m_{\rm A})$.  Since the final state in Eq.~\eqref{eq:2ParEvap} contains those amplitudes in a one-to-one map, the information about the initial state can always be found in the final state of the black hole radiation. An alternative argument to reach the same conclusion is that the explicit form of $V_{\rm evap}$ guarantees information preservation since it maps orthonormal bases into orthonormal bases, which makes it an isometry, by inspection.

\section*{Acknowledgments}
The authors would like to thank, ordered alphabetically by first name,  Andrea Sanna, Carlo Rovelli, Eugenio Bianchi, Giulia Mazzolla, Ladina Hausmann, Mariano Cadoni, Martin Renner, Mauro Oi, Nicetu Tibau Vidal, Renato Renner, Vlatko Vedral, and Xi Tong for helpful and enlightening discussions. 

This work has been supported by the Chinese Ministry of Science and Technology through grant 2023ZD0300600, by the Hong Kong Research Grant Council through grants 17307520 and SRFS2021-7S02, and by the John Templeton Foundation through grant 62312, The Quantum Information Structure of Spacetime (qiss.fr). The opinions expressed in this publication are those of the authors and do not necessarily reflect the views of the John Templeton Foundation. This research was supported in whole or in part by grants from the Austrian Science Fund (FWF) No. 10.55776/F71 and 10.55776/COE1. Research at the Perimeter Institute is supported by the Government of Canada through the Department of Innovation, Science and Economic Development Canada and by the Province of Ontario through the Ministry of Research, Innovation and Science. We acknowledge support from The City University of Hong Kong (Project No. 9610623).

\newpage

\onecolumngrid
\appendix
\setcounter{secnumdepth}{2}

\section{Unitarity of pair production}
\label{sec:squeezed}

We model the pair production as a unitary evolution.
In Ref.~\cite{Fabbri}, the operator for Hawking particle production is 
\be
\ex^{\ex^{-M k} a^\dagger_{- k} b^\dagger _k }, 
\ee
which is {\em a priori} not a unitary evolution. We demonstrate here that it is the two mode squeezing operator as it acts on the vacuum,
\be  \label{eq:squeezed}
S(\zeta)= \ex^{\zeta \left( a^\dagger_{- k} b^\dagger _k  - a_{- k} b_k \right) },
\ee
where $\zeta$ is taken to be real.

The exponent in Eq.\ref{eq:squeezed} is by inspection anti-Hermitian,  making the exponential unitary. Moreover, the squeezing operator indeed produces the Hawking pair state. Using Operator Ordering Theorems as in eqs. (A5.18-A5.25) of Ref.~\cite{QOpticsBook}, the operator ordering theorem states that, with the identification
\begin{align}
    K_+ = K_-^\dagger = a^\dagger b^\dagger \, \text{ and} \\
    K_3 = \frac{1}{2} (a^\dagger a + b b^\dagger),
\end{align}
(here we used the notation $a = a_{-k}$ and $b = a_k$), we have
\begin{equation}
    \exp(\gamma_+ K_+ + \gamma_- K_- + \gamma_3 K_3) = \exp(\Gamma_+ K_+) \exp( (\ln \Gamma_3 )K_3) \exp(  \Gamma_- K_-) 
\end{equation}
where 
\begin{align}
    \Gamma_3 = \left( \cosh \beta - \frac{\gamma_3}{2\beta }\sinh \beta \right)^{-2} \, ,\\
    \Gamma_\pm =\frac{2 \gamma_\pm \sinh \beta }{2\beta \cosh \beta - \gamma_3 \sinh \beta } \, ,
\end{align}
and
\begin{equation}
    \beta^2 = \frac{1}{4} \gamma_3 ^2 - \gamma_+ \gamma_-.
\end{equation}
In our case, $\gamma_+ = - \gamma_- = \zeta$ and $\gamma_3 = 0$, we have
\begin{equation}
    \beta = \pm \zeta
\end{equation}
and therefore
\begin{align}
    \Gamma_3 = \left( \cosh \zeta   \right)^{-2} \, , \\
    \Gamma_\pm =\frac{ \pm 2 \zeta \sinh \zeta }{2\zeta \cosh \zeta } = \pm \tanh \zeta\, ,
\end{align}
independently of the choice of $\beta= \zeta$ or $\beta = -\zeta$.

In conclusion,
\begin{align}
S(\zeta) = \exp(\zeta \left( a^\dagger b^\dagger   - a b \right) ) &= \exp((\tanh \zeta) a^\dagger b^\dagger) \exp( \frac{1}{2}(\ln (\cosh^{-2} \zeta) )  (a^\dagger a + b b^\dagger) ) \exp(  -(\tanh \zeta) ab ) = \\
&= {\rm e}^{(\tanh \zeta) a^\dagger b^\dagger} {\rm e}^{ -(\ln (\cosh \zeta) )  (a^\dagger a + b b^\dagger) } {\rm e}^{  -(\tanh \zeta) ab } ,
\end{align} 
generating a state of the form
\be S(\zeta) \ket{0,0} = \frac{1}{\cosh{\zeta}} 
\sum_{k} \left(-\tanh{\zeta}\right)^k  \ket{-k,k}.  
\ee
Following the previous comment, 
\begin{align}
\label{eq:final-state-squeezing}
    S(\zeta) \ket{0,0} =  \frac{1}{\cosh{\zeta}} \sum_{k =0}^\infty \left(\tanh{\zeta}\right)^k  \ket{-k,k}.  
\end{align}
This exactly yields the Hawking state, for $$ \zeta = - {\rm arctanh} \left( { \ex^{-M} }\right). $$ 
Since the squeezing operator is unitary, the evolution from the vacuum to the Hawking state is unitary.

\section{ The evaporation channel is approximately an isometry} \label{app:isometry}

Let $\tV^{(k)}_{\rm ev}:=V^{(k)}_{\rm ev} \cdot {V^{(k)}}^\dagger_{\rm ev}$ be the isometric channel as defined in the main text. Also, let $\map D^{(k)}:= \Tr_{{\rm rad}_k} \left[ \tV^{(k)}_{\rm ev} (\cdot)\right]$ and $\map C^{(k)}_{\rm eff}:= \Tr_{\rm bh} \left[ \tV^{(k)}_{\rm ev} (\cdot)\right]$ be the channels that are the marginals of $\tV^{(k)}_{\rm ev}$ on the black hole and radiation space, respectively. Our goal is to prove that, as $k$ goes to infinity, the marginal channel $\map C^{(k)}_{\rm eff}$ approximates better and better an isometry (approximation with respect to the diamond norm). More precisely, we will show that there exists an isometry $\map W$ such that
\begin{equation}
    \| \map C^{(k)}_{\rm eff} - \map W \|_\diamond \le 2 \, \sqrt{2\epsilon_k} \,
\end{equation}
where $\epsilon_k =  \left[ 1 -{\rm e}^{-2\pi M^2}   \left(1-  {\rm e}^{-2\pi m_* M}\right)  \right]^k$.

A key result that we will use in our derivation is a consequence of the continuity of Stinespring's dilation, we report it here for the reader's convenience.
\begin{theo}[Theorem~1 of Ref.~\cite{kretschmann2006information}]
\label{thm:continuity-of-dilation}
    Let $\tH_{\rm A}$ and $\tH_{\rm B}$ be finite-dimensional Hilbert spaces, and suppose that
    \begin{equation}
        \map T_1, \map T_2 : \map B(\tH_{\rm A}) \to \map B(\tH_{\rm B})
    \end{equation}
    are quantum channels with Stinespring isometries $V_1,\, V_2: \tH_{\rm A} \to \tH_{\rm B} \otimes \tH_{\rm E }$ and a common dilation space $\tH_{\rm E}$. Where $\mathcal B$ stands for bounded operators on the Hilbert space, and a dilation space is a space for which there exists an isometric extension.
    We then have
    \begin{equation}
        \inf_{U} \| (I_{\rm B} \otimes U) V_1 - V_2 \|^2_\infty \le \|\map T_1 - \map T_2\|_\diamond \le 2 \inf_{U} \| (I_{\rm B} \otimes U) V_1 - V_2 \|_\infty \, ,
    \end{equation}
    where the minimization is with respect to all unitaries $U \in \map B(\tH_{\rm E})$.
\end{theo}

Let $\map E : \map B(\tH_{\rm bh}) \to  \map B(\tH_{\rm bh})$ be a constant channel defined by the relation $\map E (\rho) = |0\>\<0| \Tr\left[ \rho \right]$ for any $\rho \in \map B(\tH_{\rm bh})$. Our first step in the proof is to show that 
\begin{equation}
    \| \map D^{(k)} - \map E \|_\diamond \le 2 \epsilon_k \, .
\end{equation}
For an arbitrary ancillary system E, a general pure input state has the form 
$$|\psi\>_{{\rm bh}\otimes{\rm E}} = \sum_{M=1}^{\Omega} \sum_{N=1}^{\Omega'} \alpha_{MN} \, |M\>_{\rm bh} |N\>_{\rm E} \in \tH_{\rm bh} \otimes \tH_{\rm E} \, .$$ Then, $(\tV_{\rm ev}^{(k)} \otimes \tI_{\rm E}) \, |\psi\>\<\psi| = |\psi'\>\<\psi'|$, where, from Eq.~\eqref{psik},
\begin{equation}
\label{eq:appendix-evolution-V}
    \begin{aligned}
        |\psi'\> &= \sum_{M,N} \alpha_{MN} (V^{(k)}_{\rm ev} |M\>_{\rm bh} ) |N\>_{\rm E} =\\
        &= \sum_{M,N} \sum_{m=1}^{M} \alpha_{MN} f_M^{(k)}(m) |m\>_{\rm bh} |\Lambda_{M-m}^{(k)}\>_{{\rm rad}_k}  |N\>_{\rm E} \, .
    \end{aligned}
\end{equation}
By performing the partial trace with a canonical basis on ${\rm rad}_k$, it is not too difficult to see that the resulting state will be a mixture of states in a superposition of those vectors in Eq.~\eqref{eq:appendix-evolution-V} that share the same value of $M-m$:
\begin{equation}
    \begin{aligned}
        \map D^{(k)} \otimes \map I_{\rm E} (|\psi\>\<\psi|) &= \Tr_{{\rm rad}_k} \left[ |\psi'\>\<\psi'| \right] = \\
        &= \sum_{\Delta=0}^\Omega \left( \sum_{M=\Delta}^\Omega \sum_{N=\Delta}^{\Omega'} \alpha_{MN} f_M^{(k)} (M-\Delta) |M-\Delta\> |N\>  \right) \left( \sum_{M'=\Delta}^\Omega \sum_{N'=\Delta}^{\Omega'} \alpha_{M'N'}^* {f_{M'}^{(k)}}^* (M'-\Delta) \<M'-\Delta| \<N'|  \right) \, .
    \end{aligned}
\end{equation}
On the other hand, 
\begin{equation}
    \begin{aligned}
        (\map E \otimes \map I_{\rm E}) |\psi\>\<\psi| &= |0\>\<0|_{\rm bh} \otimes \Tr_{\rm bh} \left[ |\psi\>\<\psi| \right] = \\
        &= \sum_{M = 0}^{\Omega} \sum_{N = 0}^{\Omega'} \alpha_{MN}\alpha_{MN'}^*|0\>\<0| \otimes |N\>\<N'| \, .
    \end{aligned}
\end{equation}
Then,
\begin{equation}
    \left\| \map D^{(k)} \otimes \map I_{\rm E} |\psi\>\<\psi|- \map E \otimes \map I_{\rm E} |\psi\>\<\psi|  \right\|_1 =  \sum_{\Delta=0}^\Omega \sum_{M=\Delta+1}^\Omega \sum_{N=0}^{\Omega'} |\alpha_{MN}|^2 |f_M^{(k)} (M-\Delta)|^2  + \sum_{M=0}^\Omega \sum_{N=0}^{\Omega'} |\alpha_{MN}|^2 (1- |f_M^{(k)} (0)|^2) \, .
\end{equation}
By Eq.~\eqref{exponential}, 
\begin{equation}
    \begin{aligned}
        1- |f_M^{(k)} (0)|^2 \le  \left[ 1 -{\rm e}^{-2\pi M^2}   \left(1-  {\rm e}^{-2\pi M m^*}\right)  \right]^k \,
    \end{aligned}
\end{equation}
where $m_*$ is the unit of mass. This implies that, for any constant $\alpha \ge (m^*)^2$,
\begin{equation}
    \begin{aligned}
        \sum_{M=0}^\Omega \sum_{N=0}^{\Omega'} |\alpha_{MN}|^2 (1- |f_M^{(k)} (0)|^2) &\le \sum_{M=0}^\Omega \sum_{N=0}^{\Omega'} |\alpha_{MN}|^2  \left[ 1 -{\rm e}^{-2\pi M^2}   \left(1-  {\rm e}^{-2\pi M m^*}\right)  \right]^k \le \\
        & \le \sum_{M=0}^\Omega \sum_{N=0}^{\Omega'} |\alpha_{MN}|^2  \left[ 1 -{\rm e}^{-2\pi \alpha d_{\rm bh}^2 }  \left(1-  {\rm e}^{-2\pi \alpha d_{\rm bh} }\right)  \right]^k \equiv \\
        &\equiv \sum_{M=0}^\Omega \sum_{N=0}^{\Omega'} |\alpha_{MN}|^2 \epsilon_k = \epsilon_k \, .
    \end{aligned}
\end{equation}
Concurrently, from the normalization condition, for every $\Delta \in \left[0,\Omega\right]$, $\sum_{M=\Delta+1}^\Omega |f_M^{(k)} (M-\Delta)|^2 \le  (1- |f_M^{(k)} (0)|^2) \le \epsilon_k$. In conclusion,
\begin{equation}
    \| \map D^{(k)} - \map E \|_\diamond \le 2 \epsilon_k \, .
\end{equation}

Now, let $\map E_{\rm ext} = E_{\rm ext} \cdot E_{\rm ext}^\dagger$ be the isometric extension of $\map E$ on $\map B (\tH_{\rm bh} \otimes \tH_{{\rm rad}_k})$. By applying the first inequality of Theorem~\ref{thm:continuity-of-dilation},
\begin{equation}
    \begin{aligned}
        \inf_{U} \| (I_{\rm bh} \otimes U) E_{\rm ext} - V_{\rm ev}^{(k)} \|^2_\infty &\le \|\map E - \map D^{(k)}\|_\diamond \\
        \| (I_{\rm bh} \otimes \tilde{U}) E_{\rm ext} - V_{\rm ev}^{(k)} \|_\infty &\le \sqrt{2 \epsilon_k} \,
    \end{aligned}
\end{equation} 
for some unitary $\tilde{U} \in \map B(\tH_{{\rm rad}_k})$. Let $\map E'_{\rm ext} = E'_{\rm ext} \cdot {E'_{\rm ext}}^\dagger$ for $E'_{\rm ext}:= (I_{\rm bh} \otimes \tilde{U}) E_{\rm ext}$. Since the marginal of $\map E'_{\rm ext}$ on $\tH_{\rm bh}$ is the same constant channel $\map E$, the marginal of $\map E'_{\rm ext}$ on $\tH_{{\rm rad}_k}$ must be an isometry: $\map W := \Tr_{{\rm rad}_k}\left[\map E'_{\rm ext}(\cdot)\right]$. By using the other inequality of Theorem~\ref{thm:continuity-of-dilation}, this time with dilation space $\tH_{\rm bh}$, we obtain
\begin{equation}
    \begin{aligned}
        \|\map C_{\rm eff}^{(k)} - \map W \|_\diamond \le 2 \inf_{U} \| (U \otimes I_{{\rm rad}_k}) E'_{\rm ext} - V_{\rm ev}^{(k)} \|_\infty \le 2 \| E'_{\rm ext} - V_{\rm ev}^{(k)} \|_\infty \le 2\sqrt{2} \, \sqrt{\epsilon_k} \xrightarrow{k\to\infty} 0 \, .
    \end{aligned}
\end{equation}

\section{The initial black hole state with quantum superposition}
Here we consider the case where the black hole initial state has a quantum superposition of masses. 
\be |\psi\>_{\rm bh}  =    \sum_{M=0}^{\bar M}   g(M)\,  |M\>_{\rm bh}, 
\ee for which the information is encoded in the complex amplitudes $\{g(M)\}$. 
Now we apply the evaporation map for the first time, 
\be V_{\rm ev}\ket{\psi} _{\rm bh} && =
\sum_{M}^{\bar M} g(M)  V_{\rm ev} |M\>_{\rm bh} \nonumber \\ 
&& =  \sum_{M= 0}^{\bar M} g(M)  c_M \sum_{\omega=0}^{M}   {\rm e}^{-  \pi M \omega }  |M  -\omega \>_{\rm bh}   \ket{ \omega} _{\rm out_1}  \, .
 \ee
 From Eq.\eqref{psik} we see that after k evaporation steps, 
 \begin{align}
|\psi^{(k)}\>_{{\rm bh, rad} _k}    = \sum_{M= 0}^{\bar M} g(M)
\sum_{m=0}^M  \,  f^{(k)}_M (m) ~ \,  |m\>_{\rm bh}  |\Lambda^{(k)}_{M-m}\>_{\rm rad_k}  \,. 
\end{align}
Where $|\Lambda^{(k)}_{M-m}\>_{\rm rad_k}$ is the state of $k$ radiation bursts and $f^{(k)}_M (m),  $ are the relative amplitudes.
Finally, from Eq.\eqref{exponential}, as $k$ goes to infinity, $|f^{(k\to\infty)}_m(0)| = 1 $.  Therefore, as $k\to\infty$, $|0\>_{\rm bh}$ has a probability $p_{\rm ev}=1$. Tracing the black hole state out we are then left with 
\be |\psi^{(k)}\>_{{\rm bh, rad} _k}    = \sum_{M= 0}^{\bar M} g(M)  \,  f^{(k)}_M (0) ~ \, |0\>_{\rm bh} |\Lambda^{(k)}_{M}\>_{\rm rad_k}.
\ee
It is therefore clear that the relative phases $g(M)$ are recovered in the radiation state.

\section{Information in the radiation, during the evaporation process}

We will consider one emission, from a black hole originally in the state 
\begin{align}
\left| \text{Bh}  \right\rangle = 
\sum_{m=0}^{M} f(m)  
    &   \ket{m }_{\rm A} \ket{M - m}_{\rm B} 
  \ket{0}_{\rm out}.
\end{align}

The emission, as we modelled it in the case of a two-subsystem black hole, includes a part that interacts and annihilates with system A and another one that annihilates system B. This can be expressed as

\begin{align}
\left| \text{Bh} + \text{rad} \right\rangle = 
\sum_{m=0}^{M} f(m)  
    &  \sum_{k_{\rm A}= 0}^{m} \sum_{k_{\rm B} = 0 }^{M-m} q(k_{\rm A},k_{\rm B}) \sqrt{p(k_{\rm A} + k_{\rm B})} \ket{m - k_{\rm A}}_{\rm A} \ket{M - m - k_{\rm B}}_{\rm B} 
  \ket{k_{\rm A}, k_{\rm B} }_{\rm out}.
\end{align}

Then the matrix reads,

\be
\rho && = \left| \text{Bh} + \text{rad} \right\rangle \left\langle \text{Bh} + \text{rad} \right| \\ 
&& = \sum_{m=0}^{M}\sum_{m'=0}^{M}  f^*(m') f(m)       \sum_{k_{\rm A}= 0}^{m} \sum_{k_{\rm B} = 0 }^{M-m} \sum_{l_{\rm A}=0}^{m'} \sum_{l_{\rm B}=0}^{M-m'} q^*(l_{\rm A},l_{\rm B}) q(k_{\rm A},k_{\rm B}) \sqrt{p(l_{\rm A} + l_{\rm B})p(k_{\rm A} + k_{\rm B})} \ket{m - k_{\rm A}}\bra{m' - l_{\rm A}}_{\rm A} \nonumber \\
  && \hspace{6cm} \otimes \ket{M - m - k_{\rm B}}\bra{M - m' - l_{\rm B}}_{\rm B} \otimes \ket{k_{\rm A}, k_{\rm B} } \bra{l_{\rm A}, l_{\rm B} }_{\rm out}.
\ee

Then, once we trace out the matter Hilbert space,
\be
\rho_{\rm rad} && = \Tr_{\rm bh} \{ \left| \text{Bh} + \text{rad} \right\rangle \left\langle \text{Bh} + \text{rad}  \right| \} \\ 
&& = \sum_{m=0}^{M}\sum_{m'=0}^{M}  f^*(m') f(m)       \sum_{k_{\rm A}= 0}^{m} \sum_{k_{\rm B} = 0 }^{M-m} \sum_{l_{\rm A}=0}^{m'} \sum_{l_{\rm B}=0}^{M-m'} q^*(l_{\rm A},l_{\rm B}) q(k_{\rm A},k_{\rm B}) \sqrt{p(l_{\rm A} + l_{\rm B})p(k_{\rm A} + k_{\rm B})}   \nonumber \\
  && \hspace{1cm}  \bra{m' - l_{\rm A}}\ket{m - k_{\rm A}} _{\rm A} \bra{M - m' - l_{\rm B}} \ket{M - m - k_{\rm B}}_{\rm B}  \ket{k_{\rm A}, k_{\rm B} } \bra{l_{\rm A}, l_{\rm B} }_{\rm out}.
\ee

Once we execute the sum over $l_{\rm A}$ and $l_{\rm B}$ we get,
\be
\rho_{\rm rad} &&  = \sum_{m=0}^{M}\sum_{m'=0}^{M} f^*(m') f(m)  \sum_{k_{\rm A}= 0}^{m} \sum_{k_{\rm B} = 0 }^{M-m}  q^*(k_{\rm A} + m'-m, k_{\rm B}+ m - m') q(k_{\rm A},k_{\rm B}) p(k_{\rm A} + k_{\rm B}) \nonumber \\ 
&& \hspace{6cm}   \ket{k_{\rm A}, k_{\rm B} } \bra{k_{\rm A} + m'-m, k_{\rm B}+ m - m' }_{\rm out} 
\ee

\section{ $V_{\rm ev}$ is an elementary operation} \label{app:Signalling}
Consider the case of a single degree of freedom black hole initial state in a quantum superposition of masses. 
\be |\psi\>_{\rm bh}  =    \sum_{M=0}^{\bar M}   g(M)\,  |M\>_{\rm bh}, 
\ee for which the information encoded in the complex amplitudes $\{g(M)\}$. 
Now we apply the evaporation map for the first time, 
\be \ket{\psi'} = V_{\rm ev}\ket{\psi} _{\rm bh} && =
\sum_{M}^{\bar M} g(M)  V_{\rm ev} |M\>_{\rm bh} \nonumber \\ 
&& =  \sum_{M= 0}^{\bar M} g(M)  c_M \sum_{\omega=0}^{M}   {\rm e}^{-  \pi M \omega }  |M  -\omega \>_{\rm bh}   \ket{ \omega} _{\rm out}  \, .
 \ee
We now construct the density matrix
\be \rho && = \ket{\psi'}\bra{\psi'} \\
&& =  \sum_{M'= 0}^{\bar M} \sum_{M= 0}^{\bar M} |c_M|^2 g(M) g^*(M')    \sum_{\omega=0}^{M}  \sum_{\omega'=0}^{M'}  {\rm e}^{-  \pi( M \omega +  M' \omega') } |M  -\omega \> \< M'  -\omega' |_{\rm bh}   \ket{ \omega} \bra{ \omega'} _{\rm out}  \, .
 \ee
 Tracing out the interior,
 \be \rho_{\rm out} && = \Tr_{\rm bh} [ \rho ] \\ 
 && = \sum_{M'= 0}^{\bar M} \sum_{M= 0}^{\bar M} |c_M|^2 g(M) g^*(M')    \sum_{\omega=0}^{M}  \sum_{\omega'=0}^{M'}  {\rm e}^{-  \pi( M \omega +  M' \omega') }  \< M'  - \omega' | M  -\omega \>_{\rm bh}   \ket{ \omega} \bra{ \omega'} _{\rm out}  \\
  && = \sum_{M'= 0}^{\bar M} \sum_{M= 0}^{\bar M} |c_M|^2 g(M) g^*(M')    \sum_{\omega=0}^{M}  {\rm e}^{-  \pi( M \omega +  M' (\omega + M' - M) ) }    \ket{ \omega} \bra{ \omega + M' - M} _{\rm out}  \, .
 \ee
The last equality above was reached by just evaluating the sum over $\omega'$ and implementing the Kronecker delta that arises from the inner product.

 Consider now the case of a hypothetical agent, called Alice, who can modify the state of the black hole before the evaporation operation is applied. We assume that she can do so by applying a general unitary transformation $U$. In other words, she can modify the state into 
\be |\tilde \psi\>_{\rm bh}  =    \sum_{M=0}^{\bar M}   g(M)\,  U |M\>_{\rm bh} = \sum_{M=0}^{\bar M}   \tilde g(M)\,   |M\>_{\rm bh},
\ee
 which would upon annihilation yield outside, 
\be \tilde \rho_{\rm out} && = \sum_{M'= 0}^{\bar M} \sum_{M= 0}^{\bar M} |c_M|^2 \tilde{g}(M) \tilde{g}^*(M')    \sum_{\omega=0}^{M}  {\rm e}^{-  \pi( M \omega +  M' (\omega + M' - M) ) }    \ket{ \omega} \bra{ \omega + M' - M} _{\rm out}  \, .
 \ee

Since the \( \tilde{g} \)'s carry information about the encoded unitary, this information can, in principle, be observed from the outside—say, by an agent Bob. This is not surprising for two reasons. First, \( V_{\rm ev} \) is a global unitary (or more generally, a channel from Alice to Bob), and as such, it is capable of affecting the outgoing radiation and imprinting black hole information onto it.
Second, this signaling respects relativistic causality in the following sense. In agreement with standard black hole physics, we assume that total black hole mass information is encoded at the horizon. Therefore, Alice can modify the black hole state before the evaporation operation \( V_{\rm ev} \) is applied, but the resulting information is still encoded at the horizon. Since the black hole information, upon application of the evaporation operation, is transferred to the outgoing modes of Hawking radiation—and this radiation must propagate from the horizon to Bob’s location—there is no faster-than-light signaling from Alice to Bob in this case.

This discussion applies to the scenario in which we treat \( V_{\rm ev} \) as a single elementary operation, rather than splitting it into separate components. The relevance of this assumption will become apparent in the discussion that follows.
\color{black}

If we split \( V_{\rm ev} \) into its components—the particle production \( V_{\rm prod} \) and the annihilation \( W \)—and treat these as two separate operations between which Alice can intervene by applying a unitary, then this process may lead to instantaneous signaling. For example, suppose that the state after pair production but before annihilation is 
\be 
 \ket{\phi} && = V_{\rm prod} \ket{\psi} _{\rm bh} \\ 
 && = \sum_{M}^{\bar M} g(M)  V_{\rm prod} |M\>_{\rm bh} \nonumber \\ 
&& =  \sum_{M= 0}^{\bar M} g(M)  c_M \sum_{\omega=0}^{M}   {\rm e}^{-  \pi M \omega }  |M  \>_{\rm bh} |-\omega \>_{\rm int}   \ket{ \omega} _{\rm out}  \, .
\ee
Suppose, moreover, that just before the annihilation,  Alice applies a unitary U to the state of the black hole system, resulting in 
\be U \ket{\phi}  && =  \sum_{M= 0}^{\bar M} g'(M)  c_M \sum_{\omega=0}^{M}   {\rm e}^{-  \pi M \omega }  |M  \>_{\rm bh} |-\omega \>_{\rm int}   \ket{ \omega} _{\rm out}.
\ee
Then, if the annihilation happen immediately after the unitary is applied, the effect of the unitary is instantaneously signaled to the outside observer, Bob, who has already received the outgoing Hawking radiation in the meantime and has access to the state 
\be 
 \rho'_{\rm out} && = \sum_{M'= 0}^{\bar M} \sum_{M= 0}^{\bar M} |c_M|^2 {g'}(M){g'}^*(M')    \sum_{\omega=0}^{M}  {\rm e}^{-  \pi( M \omega +  M' (\omega + M' - M) ) }    \ket{ \omega} \bra{ \omega + M' - M} _{\rm out}~.
\ee
This instantaneous signalling implies that the evaporation operator should be treated as a black box and a whole process rather than its parts (the creation and especially the annihilation) being taken seriously as fundamental processes.
This is consistent with the negative energy particles being off-shell particles, as is the case in their other occurrences in the Standard Model of particle physics \cite{Weinberg}.

\section{Details on numerical simulation}

Below is the self-contained Matlab code used to generate Figure 2.

\begin{lstlisting}
%Generating Entropy and Mass curves of unitary black hole evaporation model.
%Four systems involved: BH1, BH2, INT, OUT. Basis is {|w>} with w=0,1,2,3,4
%for each system.
%Mass of first two particles is M12=w1+w2.
%The dynamics in each step involves bringing in new radiation mode, 
%doing a controlled squeezing unitary, and an annihilation
%step. The annihilation step involves some amplitude for interacting with 
%BH1 as well as BH2, modelled via implementing a partial swap unitary on 
%BH1 and BH2 before annihilation with BH1.
clear

dt=0.1;%time interval, basically U=expm(-iHdt).

%---------------------------------
%I USEFUL SHORTHANDS

%convenient short-hand notation
ket0=[1;0;0;0;0];ket1=[0;1;0;0;0];ket2=[0;0;1;0;0];
ket3=[0;0;0;1;0];ket4=[0;0;0;0;1];

ket00=kron(ket0,ket0);ket01=kron(ket0,ket1);ket02=kron(ket0,ket2);
ket03=kron(ket0,ket3);ket04=kron(ket0,ket4);

ket10=kron(ket1,ket0);ket11=kron(ket1,ket1);ket12=kron(ket1,ket2);
ket13=kron(ket1,ket3);ket14=kron(ket1,ket4);

ket20=kron(ket2,ket0);ket21=kron(ket2,ket1);ket22=kron(ket2,ket2);
ket23=kron(ket2,ket3);ket24=kron(ket2,ket4);

ket30=kron(ket3,ket0);ket31=kron(ket3,ket1);ket32=kron(ket3,ket2);
ket33=kron(ket3,ket3);ket34=kron(ket3,ket4);

ket40=kron(ket4,ket0);ket41=kron(ket4,ket1);ket42=kron(ket4,ket2);
ket43=kron(ket4,ket3);ket44=kron(ket4,ket4);

ket110=kron(ket11,ket0);ket111=kron(ket11,ket1);ket112=kron(ket11,ket2);
ket113=kron(ket11,ket3);ket114=kron(ket11,ket4);

ket220=kron(ket22,ket0);ket221=kron(ket22,ket1);ket222=kron(ket22,ket2);
ket223=kron(ket22,ket3);ket224=kron(ket22,ket4);

ket330=kron(ket33,ket0);ket331=kron(ket33,ket1);ket332=kron(ket33,ket2);
ket333=kron(ket33,ket3);ket334=kron(ket33,ket4);

ket440=kron(ket44,ket0);ket441=kron(ket44,ket1);ket442=kron(ket44,ket2);
ket443=kron(ket44,ket3);ket444=kron(ket44,ket4);

ket000=kron(ket00,ket0);
ket001=kron(ket00,ket1);
ket002=kron(ket00,ket2);
ket003=kron(ket00,ket3);
ket004=kron(ket00,ket4);
ket100=kron(ket10,ket0);
ket101=kron(ket10,ket1);
ket102=kron(ket10,ket2);
ket103=kron(ket10,ket3);
ket104=kron(ket10,ket4);
ket200=kron(ket20,ket0);
ket201=kron(ket20,ket1);
ket202=kron(ket20,ket2);
ket203=kron(ket20,ket3);
ket204=kron(ket20,ket4);
ket300=kron(ket30,ket0);
ket301=kron(ket30,ket1);
ket302=kron(ket30,ket2);
ket303=kron(ket30,ket3);
ket304=kron(ket30,ket4);
ket401=kron(ket40,ket1);
ket402=kron(ket40,ket2);
ket403=kron(ket40,ket3);
ket404=kron(ket40,ket4);

%---------------------------------
%II CREATING SQUEEZING OPERATOR
%The squeezing operator involves creation and annihilation operators. 

%Define the annihiliation and creation operators initially on the systems
%alone via standard dirac notation expresssion for these operators
CREATE=zeros(5,5);
for ic=1:4
CREATE(ic+1,ic)=sqrt(ic);
end
ANNIH=CREATE'; %' means dagger

%creation/annih operators on OUT mode
CREATEOUT=kron(eye(5,5),CREATE);
ANNIHOUT=CREATEOUT';
%creation/annih operators on INT mode
CREATEINT=kron(CREATE,eye(5,5)); 
ANNIHINT=CREATEINT';

SQUEEZEANDEATEXPONENT=ANNIHINT*CREATEOUT-CREATEINT*ANNIHOUT;
%This exponent will be used in the unitary evolution later.

%---------------------------------
%III MODEL INTERACTION BETWEEN INT AND BH: EAT

%projectors onto BH mass subspaces
P0=kron(ket00,ket00');
P1=kron(ket01,ket01')+kron(ket10,ket10');
P2=kron(ket02,ket02')+kron(ket20,ket20')+kron(ket11,ket11');
P3=kron(ket03,ket03')+kron(ket30,ket30')+kron(ket12,ket12')...
+kron(ket21,ket21');
P4=kron(ket04,ket04')+kron(ket40,ket40')+kron(ket13,ket13')...
+kron(ket31,ket31')+kron(ket22,ket22');
P5=kron(ket14,ket14')+kron(ket41,ket41')+kron(ket23,ket23')...
+kron(ket32,ket32');
P6=kron(ket24,ket24')+kron(ket42,ket42')+kron(ket33,ket33');
P7=kron(ket43,ket43')+kron(ket34,ket34');
P8=kron(ket44,ket44');

%mass observable
M=0*P0+1*P1+2*P2+3*P3+4*P4+5*P5+6*P6+7*P7+8*P8;


%Key part of Unitary Evolution
%k=1;
fofM=expm(-M);
SQUEEZEANDEAT=kron(eye(5,5),expm(fofM*SQUEEZEANDEATEXPONENT));

%optional check of unitarity
%norm(SQUEEZEANDEAT'*SQUEEZEANDEAT-eye(5^2,5^2))


%-----------------------------------
%IV ONE MORE UNITARY. Finally, we want to be able for the int mode to eat 
%also the first particle so introduce a swapping amplitude
%between 1 and 2. 

%SWAP UNITARY
SWAP=...
 kron(ket00,ket00')+kron(ket10,ket01')+kron(ket20,ket02')+kron(ket30,ket03')...
+kron(ket40,ket04')...
+kron(ket01,ket10')+kron(ket11,ket11')+kron(ket21,ket12')+kron(ket31,ket13')...
+kron(ket41,ket14')...
+kron(ket02,ket20')+kron(ket12,ket21')+kron(ket22,ket22')+kron(ket32,ket23')...
+kron(ket42,ket24')...
+kron(ket03,ket30')+kron(ket13,ket31')+kron(ket23,ket32')+kron(ket33,ket33')...
+kron(ket43,ket34')...
+kron(ket04,ket40')+kron(ket14,ket41')+kron(ket24,ket42')+kron(ket34,ket43')...
+kron(ket44,ket44');...

SWAP12=kron(SWAP,eye(5,5));
SMOOTHSWAP12=expm(1i*0.3*SWAP12);

%-----------------------------------
%V TIME EVOLVE STATE
loops=5000


%take initial state as |x,x,0> where systems are
%|BH1,BH2,OUT>
initialstate=ket440;
state=initialstate;
rho=kron(state,state');


p0vec=zeros(loops,1);
OUTPurityVec=zeros(loops,1);
Mvec=zeros(loops,1);%to store BH mass
Mfull=kron(M, eye(5,5));%to measure BH mass

%total dynamics
UTOT=SQUEEZEANDEAT*SMOOTHSWAP12;
UTOTDAGGER=UTOT';

%Projector to be used when bringing in new systems
P0single=kron(ket0,ket0');%P0 is on 2 systems, P0single on one.

loopcount=0;
for loopcount=1:loops
    %record mass expectation value of BH1 and BH2 
    Mvec(loopcount)=trace(Mfull*rho);

    %reset OUTto 0 (like introducing new systems to interact with)
    rho12=traceoutlastsystem(rho,3);
    rho=kron(rho12,P0single);

    %unitary evolution hits them all
    rho=UTOT*rho*UTOTDAGGER;

    %check purity (tr(rho^2)) of BH
    OUTPurityVec(loopcount)=trace(rho12*rho12);

    %record expected mass 
    Mvec(loopcount)=trace(Mfull*rho); 
end


%-----------------------------------
%VI PLOT

% Plot the first curve
plot([1:loops]', abs(-log2(OUTPurityVec)), 'b-', 'LineWidth', 2); 
% 'b-' specifies blue color and solid line

% Add the second curve to the same plot
hold on; % To hold the current plot
plot([1:loops]', abs(Mvec), 'r--', 'LineWidth', 2); 
% 'r--' specifies red color and dashed line

% Customize the plot
xlabel('Time');
ylabel('Value');
title('Mass and entropy of black hole');
legend('Renyi-2 entropy of black hole', 'Mass of black hole'); 

hold off;
\end{lstlisting}

\twocolumngrid


\begin{thebibliography}{99}
\bibitem{NegativeE}
Normally, negative energy matter is not allowed by the laws of physics, however, inside the black hole, since the time dimension and the radial space dimension switch roles, what is negative energy from outside is seen inside as a negative momentum, thus not violating the laws of physics.
\end{thebibliography}
\end{document}